\documentclass[twoside]{article}
\usepackage[comma,authoryear]{natbib}
\usepackage{epsfig}
\usepackage{url}

\newlength{\figurewidth}
\renewcommand{\sectionmark}[1]{\markboth{\textsc{Leskovec \& Horvitz}}
{\textsc{Planetary-Scale IM Network}}}
\renewcommand{\subsectionmark}[1]{\markboth{\textsc{Leskovec \& Horvitz}}
{\textsc{Planetary-Scale IM Network}}}

\newcommand{\hide}[1]{}

\newcommand{\reminder}[1]{}
\renewcommand{\cite}[1]{\citep{#1}}

\newcommand{\captionfonts}{\small}
\makeatletter
\long\def\@makecaption#1#2{%
  \vskip\abovecaptionskip
  \sbox\@tempboxa{{\captionfonts #1: #2}}%
  \ifdim \wd\@tempboxa >\hsize
    {\captionfonts #1: #2\par}
  \else
    \hbox to\hsize{\hfil\box\@tempboxa\hfil}%
  \fi
  \vskip\belowcaptionskip}
\makeatother

\begin{document}

\title{Planetary-Scale Views on an Instant-Messaging Network\footnote{
Shorter version of this work appears in the WWW '08: Proceedings of the 16th
international conference on World Wide Web, 2008.} }

\author{
  Jure Leskovec\footnote{This work was performed while the first author
  was an intern at Microsoft Research.}\\
    \textit{\normalsize Machine Learning Department}\\
    \textit{\normalsize Carnegie Mellon University}\\
    \textit{\normalsize Pittsburgh, PA, USA}\\
    \\
  Eric Horvitz\\
    \textit{\normalsize Microsoft Research}\\
    \textit{\normalsize Redmond, WA, USA}\\
    \\
    \\
    Microsoft Research Technical Report\\
    MSR-TR-2006-186\\
}

\date{June 2007}

\maketitle
\begin{abstract}
We present a study of anonymized data capturing a month of high-level
communication activities within the whole of the Microsoft Messenger
instant-messaging system. We examine characteristics and patterns that emerge
from the collective dynamics of large numbers of people, rather than the
actions and characteristics of individuals. The dataset contains summary
properties of 30 billion conversations among 240 million people. From the
data, we construct a communication graph with 180 million nodes and 1.3
billion undirected edges, creating the largest social network constructed and
analyzed to date. We report on multiple aspects of the dataset and
synthesized graph. We find that the graph is well-connected and robust to
node removal.  We investigate on a planetary-scale the oft-cited report that
people are separated by ``six degrees of separation'' and find that the
average path length among Messenger users is 6.6. We also find that people
tend to communicate more with each other when they have similar age,
language, and location, and that cross-gender conversations are both more
frequent and of longer duration than conversations with the same gender.
\end{abstract}

\newpage
\section{Introduction} \label{sec:intro}
\vspace{6mm}

Large-scale web services provide unprecedented opportunities to capture and
analyze behavioral data on a planetary scale.  We discuss findings drawn from
aggregations of anonymized data representing one month (June 2006) of
high-level communication activities of people using the Microsoft Messenger
instant-messaging (IM) network. We did not have nor seek access to the
content of messages. Rather, we consider structural properties of a
communication graph and study how structure and communication relate to user
demographic attributes, such as gender, age, and location.  The data set
provides a unique lens for studying patterns of human behavior on a wide
scale.

We explore a dataset of 30 billion conversations generated by 240 million
distinct users over one month. We found that approximately 90 million
distinct Messenger accounts were accessed each day and that these users
produced about 1 billion conversations, with approximately 7 billion
exchanged messages per day. 180 million of the 240 million active accounts
had at least one conversation on the observation period. We found that 99\%
of the conversations occurred between 2 people, and the rest with greater
numbers of participants. To our knowledge, our investigation represents the
largest and most comprehensive study to date of presence and communications
in an IM system. A recent report~\cite{idc05messenger} estimated that
approximately 12 billion instant messages are sent each day.  Given the
estimate and the growth of IM, we estimate that we captured approximately
half of the world's IM communication during the observation period.

We created an undirected {\em communication network} from the data where each
user is represented by a node and an edge is placed between users if they
exchanged at least one message during the month of observation. The network
represents accounts that were active during June 2006. In summary, the
communication graph has 180 million nodes, representing users who
participated in at least one conversation, and 1.3 billion undirected edges
among active users, where an edge indicates that a pair of people
communicated. We note that this graph should be distinguished from a buddy
graph where two people are connected if they appear on each other's contact
lists. The buddy graph for the data contains 240 million nodes and 9.1
billion edges.  On average each account has approximately 50 buddies on a
contact list.

To highlight several of our key findings, we discovered that the
communication network is well connected, with 99.9\% of the nodes belonging
to the largest connected component. We evaluated the oft-cited finding by
Travers and Milgram that any two people are linked to one another on average
via a chain with ``6-degrees-of-separation'' \cite{milgram69smallworld}.  We
found that the average shortest path length in the Messenger network is 6.6
(median 6), which is half a link more than the path length measured in the
classic study. However, we also found that longer paths exist in the graph,
with lengths up to 29. We observed that the network is well clustered, with a
clustering coefficient~\cite{watts98smallworld} that decays with exponent
$-0.37$. This decay is significantly lower than the value we had expected
given prior research~\cite{ravasz03hierar}. We found strong {\em
homophily}~\cite{mcpherson01homophily,rogers70homophily} among users; people
have more conversations and converse for longer durations with people who are
similar to themselves.  We find the strongest homophily for the language
used, followed by conversants' geographic locations, and then age.  We found
that homophily does not hold for gender; people tend to converse more
frequently and with longer durations with the opposite gender. We also
examined the relation between communication and distance, and found that the
number of conversations tends to decrease with increasing geographical
distance between conversants. However, communication links spanning longer
distances tend to carry more and longer conversations.

\section{Instant Messaging} \label{sec:data}

The use of IM has been become widely adopted in personal and businesss
communications. IM clients allow users fast, near-synchronous communication,
placing it between synchronous communication mediums, such as real-time voice
interactions, and asynchronous communication mediums like
email~\cite{voida02messaging}. IM users exchange short text messages with one
or more users from their list of contacts, who have to be on-line and logged
into the IM system at the time of interaction. As conversations and messages
exchanged within them are usually very short, it has been observed that users
employ informal language, loose grammar, numerous abbreviations, with minimal
punctuation~\cite{nardi00messaging}. Contact lists are commonly referred to
as {\em buddy lists} and users on the lists are referred to as {\em buddies}.

\subsection{Research on Instant Messaging}

Several studies on smaller datasets are related to this work. Avrahami and
Hudson~\cite{avrahami06im} explored communication characteristics of 16 IM
users. Similarly, Shi et al.~\cite{shi06strongties} analyzed IM contact lists
submitted by users to a public website and explored a static contact network
of 140,000 people. Recently, Xiao et al.~\cite{xiao07traffic} investigated IM
traffic characteristics within a large organization with 400 users of
Messenger. Our study differs from the latter study in that we analyze the
{\em full} Messenger population over a one month period, capturing the
interaction of user demographic attributes, communication patterns, and
network structure.

\subsection{Data description}

To construct the Microsoft Instant Messenger communication dataset, we
combined three different sources of data: (1) user demographic information,
(2) time and user stamped events describing the presence of a particular
user, and (3) communication session logs, where, for all participants, the
number of exchanged messages and the periods of time spent participating in
sessions is recorded.

We use the terms {\em session} and {\em conversation} interchangeably to
refer to an IM interaction among two or more people. Although the Messenger
system limits the number of people communicating at the same time to 20,
people can enter and leave a conversation over time. We note that, for large
sessions, people can come and go over time, so conversations can be long with
many different people participating. We observed some very long sessions with
more than 50 participants joining over time.

All of our data was anonymized; we had no access to personally identifiable
information. Also, we had no access to text of the messages exchanged or any
other information that could be used to uniquely identify users. We focused
on analyzing high-level characteristics and patterns that emerge from the
collective dynamics of 240 million people, rather than the actions and
characteristics of individuals. The analyzed data can be split into three
parts: {\em presence data}, {\em communication data}, and {\em user
demographic information}:
\begin{itemize}
  \item {\bf Presence events:} These include login, logout, first ever
      login, add, remove and block a buddy, add unregistered buddy
      (invite new user), change of status (busy, away, be-right-back,
      idle, etc.). Events are user and time stamped.
  \item{\bf Communication:} For each user participating in the session,
      the log contains the following tuple: session id, user id, time
      joined the session, time left the session, number of messages sent,
      number of messages received.
  \item{\bf User data:} For each user, the following self-reported
      information is stored: age, gender, location (country, ZIP),
      language, and IP address. We use the IP address to decode the
      geographical coordinates, which we then use to position users on
      the globe and to calculate distances.
\end{itemize}

We gathered data for 30 days of June 2006. Each day yielded about 150
gigabytes of compressed text logs (4.5 terabytes in total). Copying the data
to a dedicated eight-processor server with 32 gigabytes of memory took 12
hours. Our log-parsing system employed a pipeline of four threads that parse
the data in parallel, collapse the session join/leave events into sets of
conversations, and save the data in a compact compressed binary format. This
process compressed the data down to 45 gigabytes per day. Processing the data
took an additional 4 to 5 hours per day.

A special challenge was to account for missing and dropped events, and
session ``id recycling" across different IM servers in a server farm.
As part of this process, we closed a session 48 hours after the last leave
session event. We closed sessions automatically if only one user was left in
the conversation.

\section{Usage \& population statistics}

We shall first review several statistics drawn from aggregations of users and
their communication activities.

\subsection{Levels of activity}

Over the observation period, 242,720,596 users logged into Messenger and
179,792,538 of these users were actively engaged in conversations by sending
or receiving at least one IM message. Over the month of observation,
17,510,905 new accounts were activated.
As a representative day, on June 1 2006, there were almost 1 billion
(982,005,323) different sessions (conversations among any number of people),
with more than 7 billion IM messages sent. Approximately 93 million users
logged in with 64 million different users becoming engaged in conversations
on that day.  Approximately 1.5 million new users that were not registered
within Microsoft Messenger were invited to join on that particular day.

\begin{figure}
\begin{center}
  \begin{tabular}{cc}
    \includegraphics[width=0.45\textwidth]{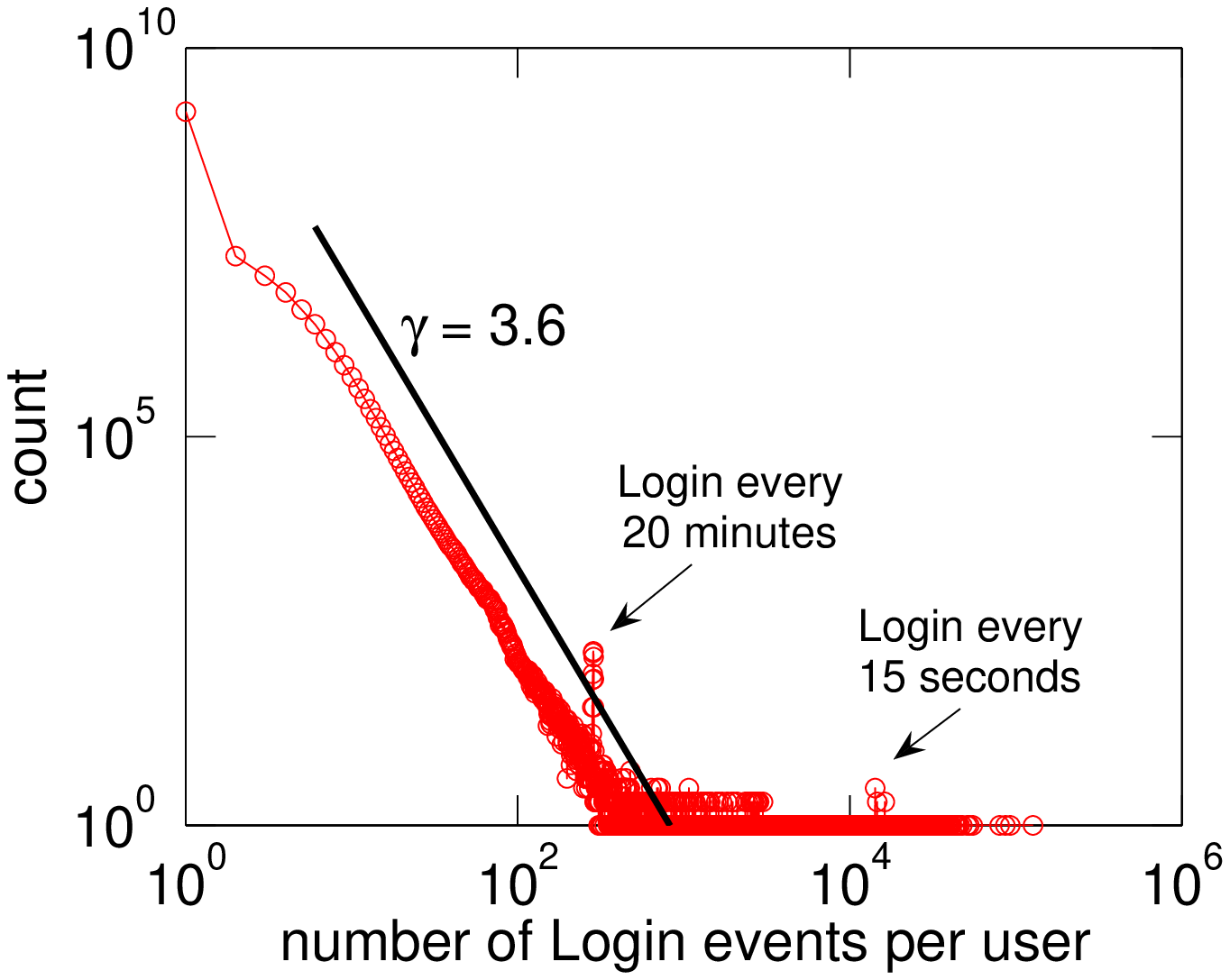} &
    \includegraphics[width=0.45\textwidth]{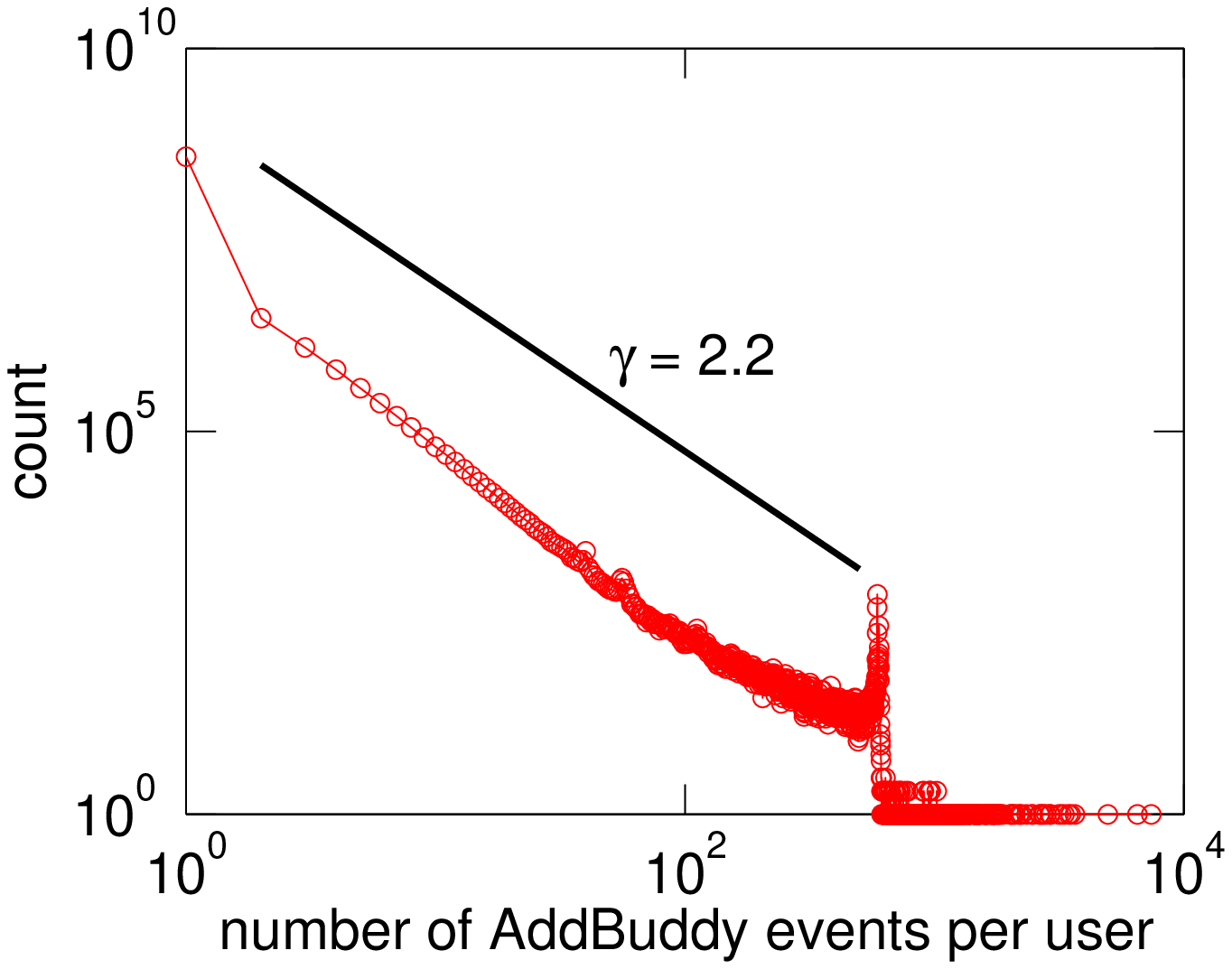} \\
    (a) Login & (b) AddBuddy \\
  \end{tabular}
  \caption{Distribution of the number of events per user.
  (a) Number of logins per user. (b) Number of
  buddies added per user.}
  \label{fig:eventsPerUsr}
\end{center}
\end{figure}

\begin{figure}
\begin{center}
    \begin{tabular}{cc}
    \includegraphics[width=0.45\textwidth]{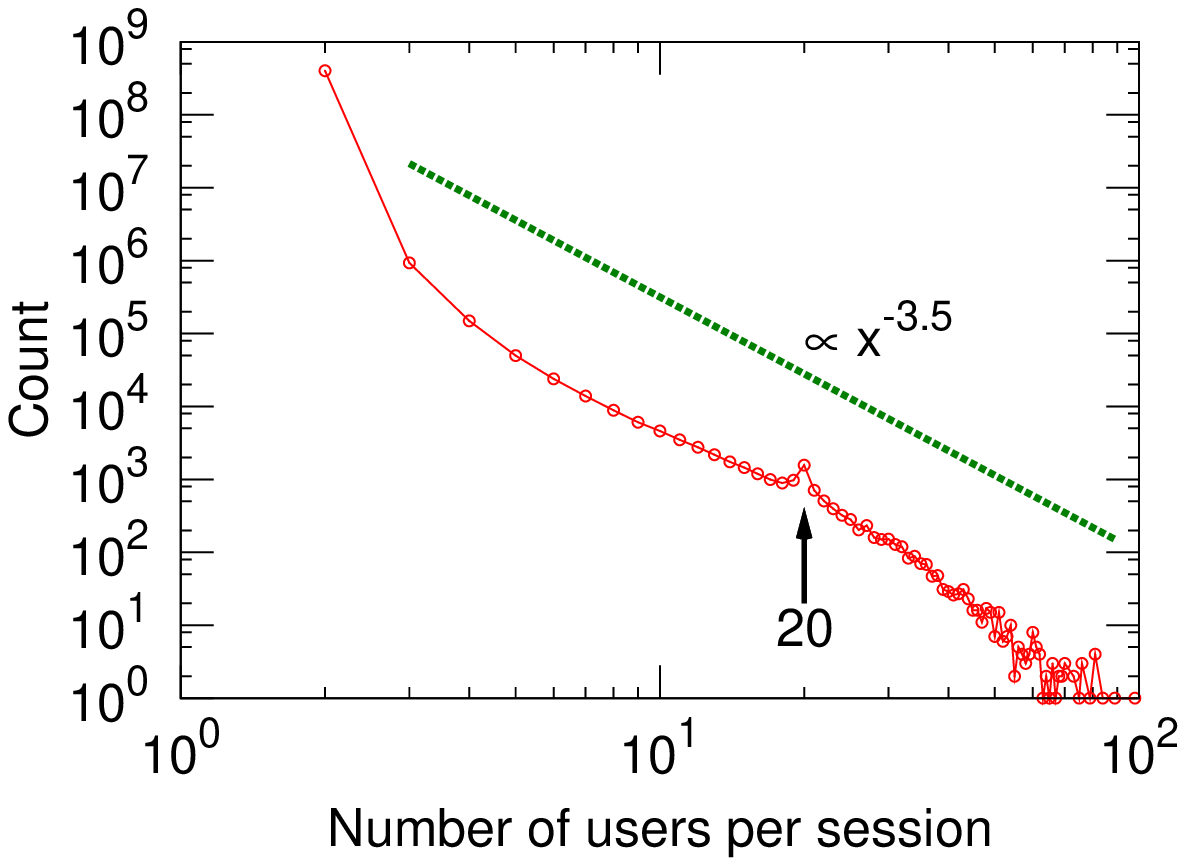} &
    \includegraphics[width=0.45\textwidth]{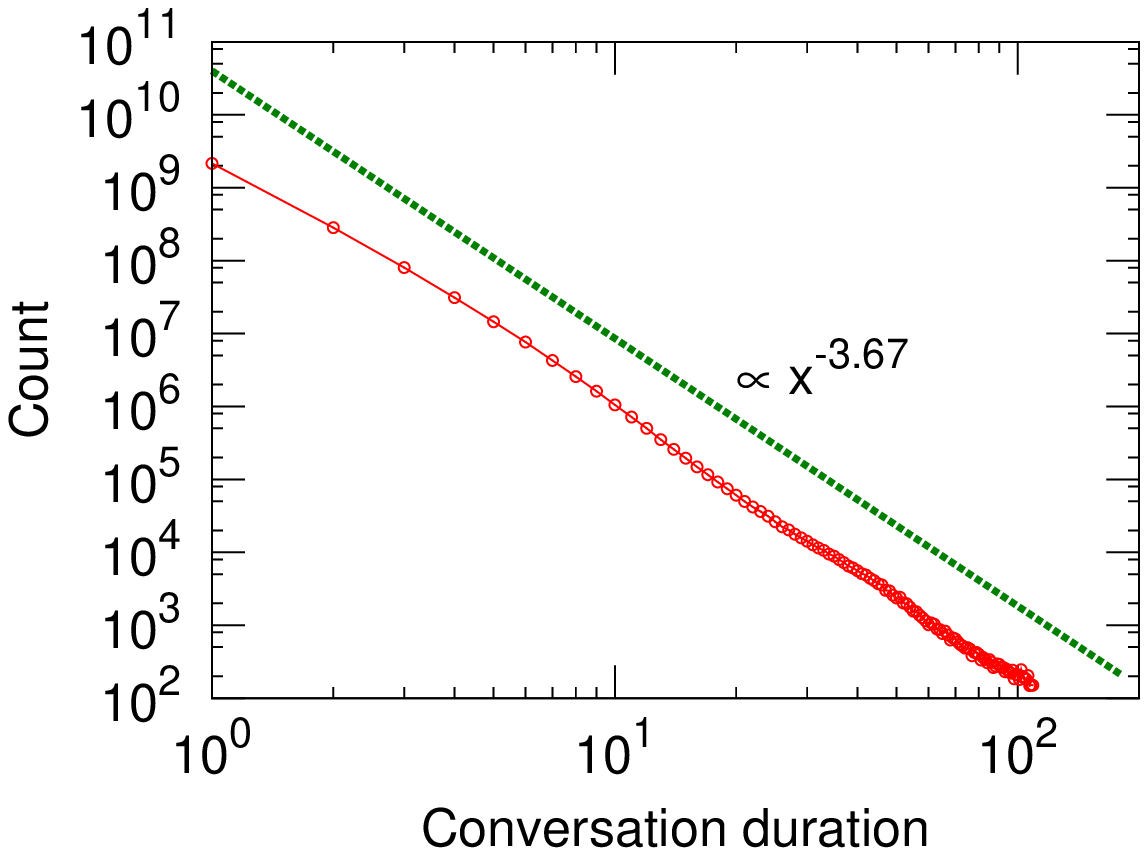} \\
    \end{tabular}
  \caption{(a) Distribution of the number of people participating in a
  conversation. (b) Distribution of the durations of conversations. The spread of durations can be described by a
  power-law distribution.}
  \label{fig:buddiesB}
  \label{fig:buddiesC}
\end{center}
\end{figure}

We consider event distributions on a per-user basis in
Figure~\ref{fig:eventsPerUsr}. The number of logins per user, displayed in
Figure~\ref{fig:eventsPerUsr}(a), follows a heavy-tailed distribution with
exponent 3.6. We note spikes in logins at 20 minute and 15 second intervals,
which correspond to an auto-login function of the IM client. As shown in
Figure~\ref{fig:eventsPerUsr}(b), many users fill up their contact lists
rather quickly. The spike at 600 buddies undoubtedly reflects the maximal
allowed length of contact lists.

Figure~\ref{fig:buddiesB}(a) displays the number of users per session. In
Messenger, multiple people can participate in conversations. We observe a
peak at 20 users, the limit on the number of people who can participate
simultaneously in a conversation. Figure~\ref{fig:buddiesB}(b) shows the
distribution over the session durations, which can be modeled by a power-law
distribution with exponent $~3.6$.

\begin{figure}[t]
\begin{center}
    \begin{tabular}{cc}
    \includegraphics[width=0.45\textwidth]{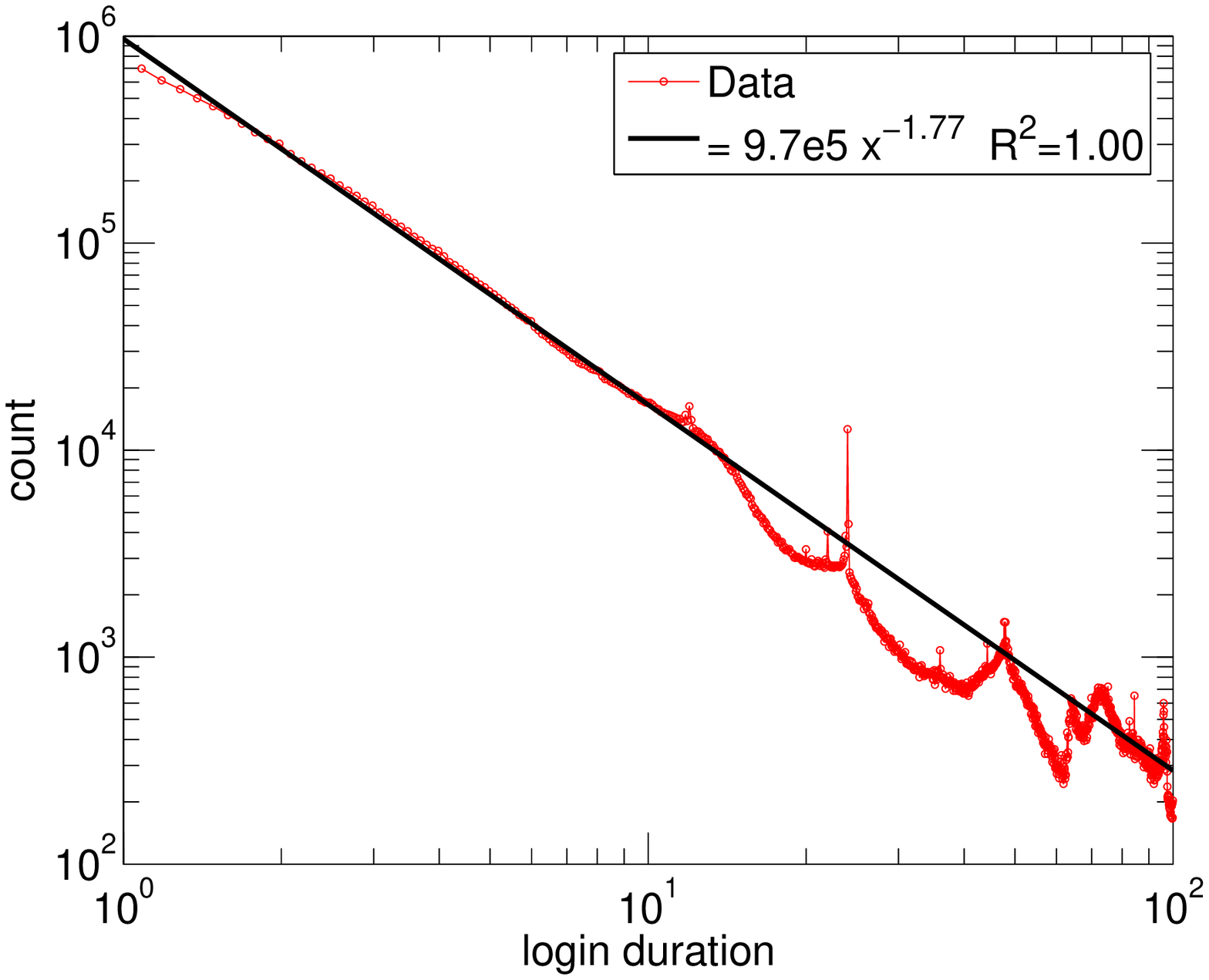} &
    \includegraphics[width=0.45\textwidth]{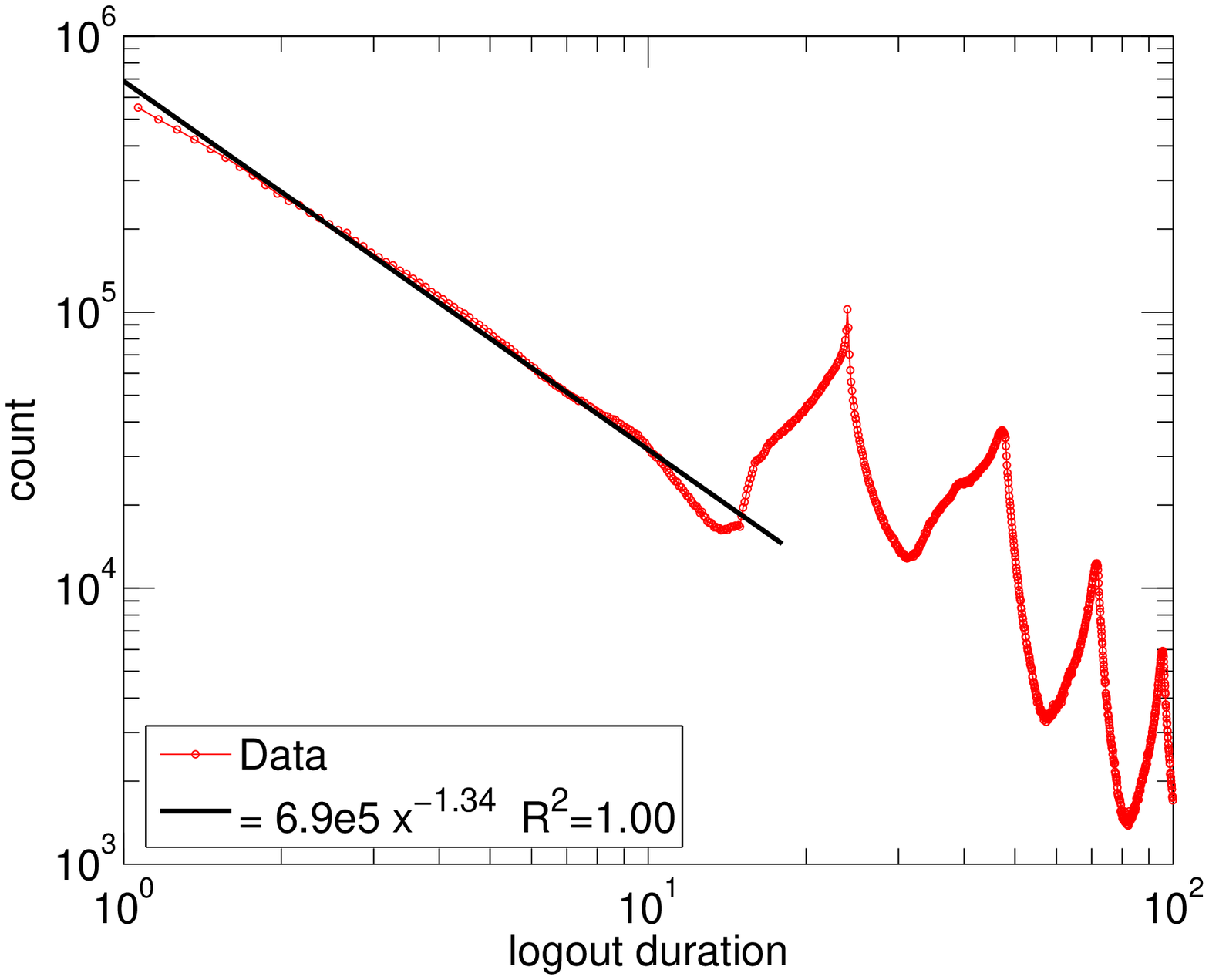} \\
    \end{tabular}
    \caption{(a) Distribution of login duration.
   (b) Duration of times when people are not logged into the system
   (times between logout and login).}
   \label{fig:loginDurationA}
   \label{fig:loginDurationB}
\end{center}
\end{figure}

Next, we examine the distribution of the durations of periods of time when
people are logged on to the system. Let $(ti_j, to_j)$ denote a time ordered
($ti_j < to_j < ti_{j+1}$) sequence of online and offline times of a user,
where $ti_j$ is the time of the $j$th login, and $to_j$ is the corresponding
logout time. Figure~\ref{fig:loginDurationA}(a) plots the distribution of
$to_j - ti_j$ over all $j$ over all users. Similarly,
Figure~\ref{fig:loginDurationB}(b) shows the distribution of the periods of
time when users are logged off, {\em i.e.} $ti_{j+1}-to_j$ over all $j$ and
over all users. Fitting the data to power-law distributions reveals exponents
of 1.77 and 1.3, respectively. The data shows that durations of being online
tend to be shorter and decay faster than durations that users are offline. We
also notice periodic effects of login durations of 12, 24, and 48 hours,
reflecting daily periodicities. We observe similar periodicities for logout
durations at multiples of 24 hours.

\begin{figure}
\begin{center}
  \begin{tabular}{ccc}
    \includegraphics[width=0.32\textwidth]{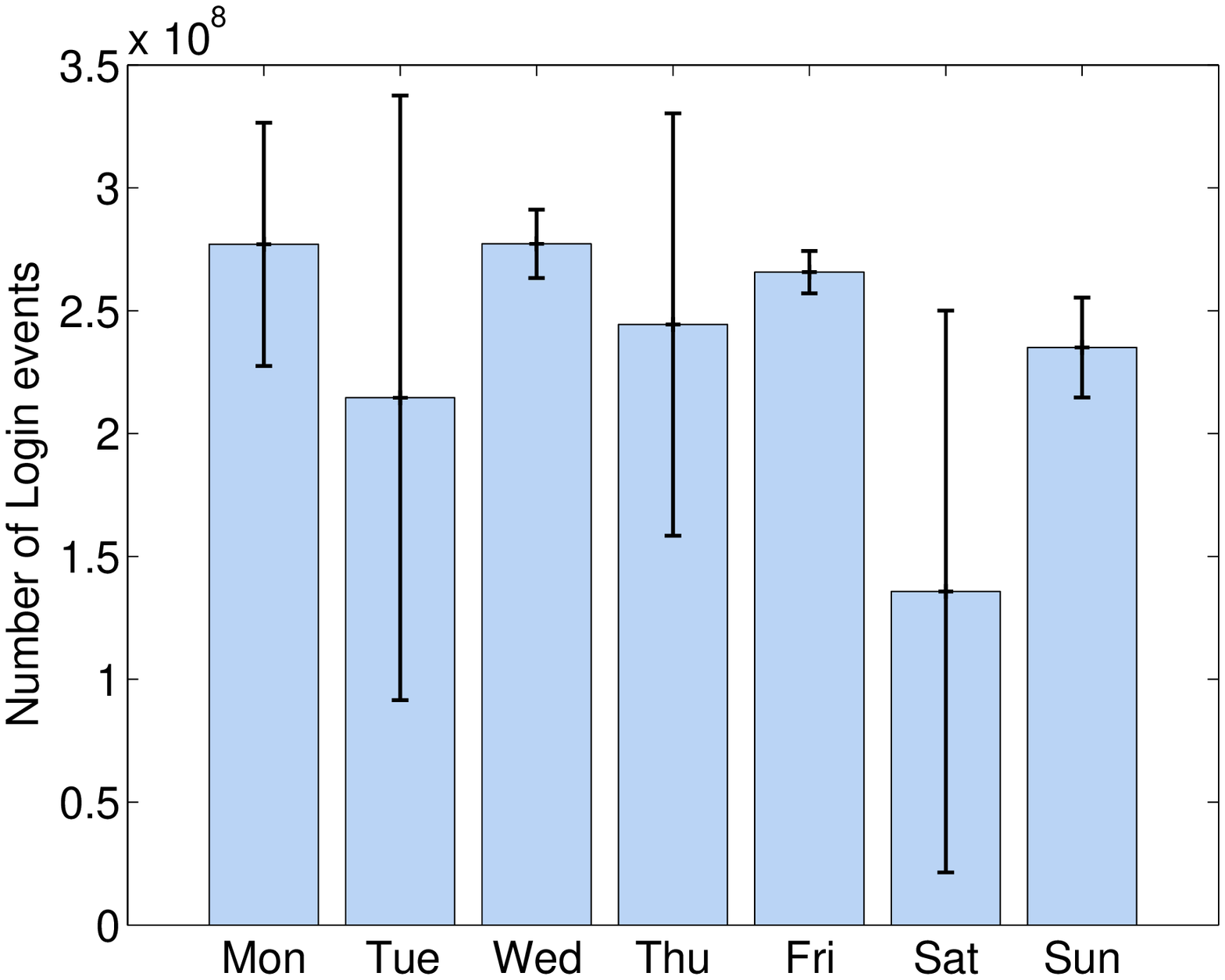} &
    \includegraphics[width=0.32\textwidth]{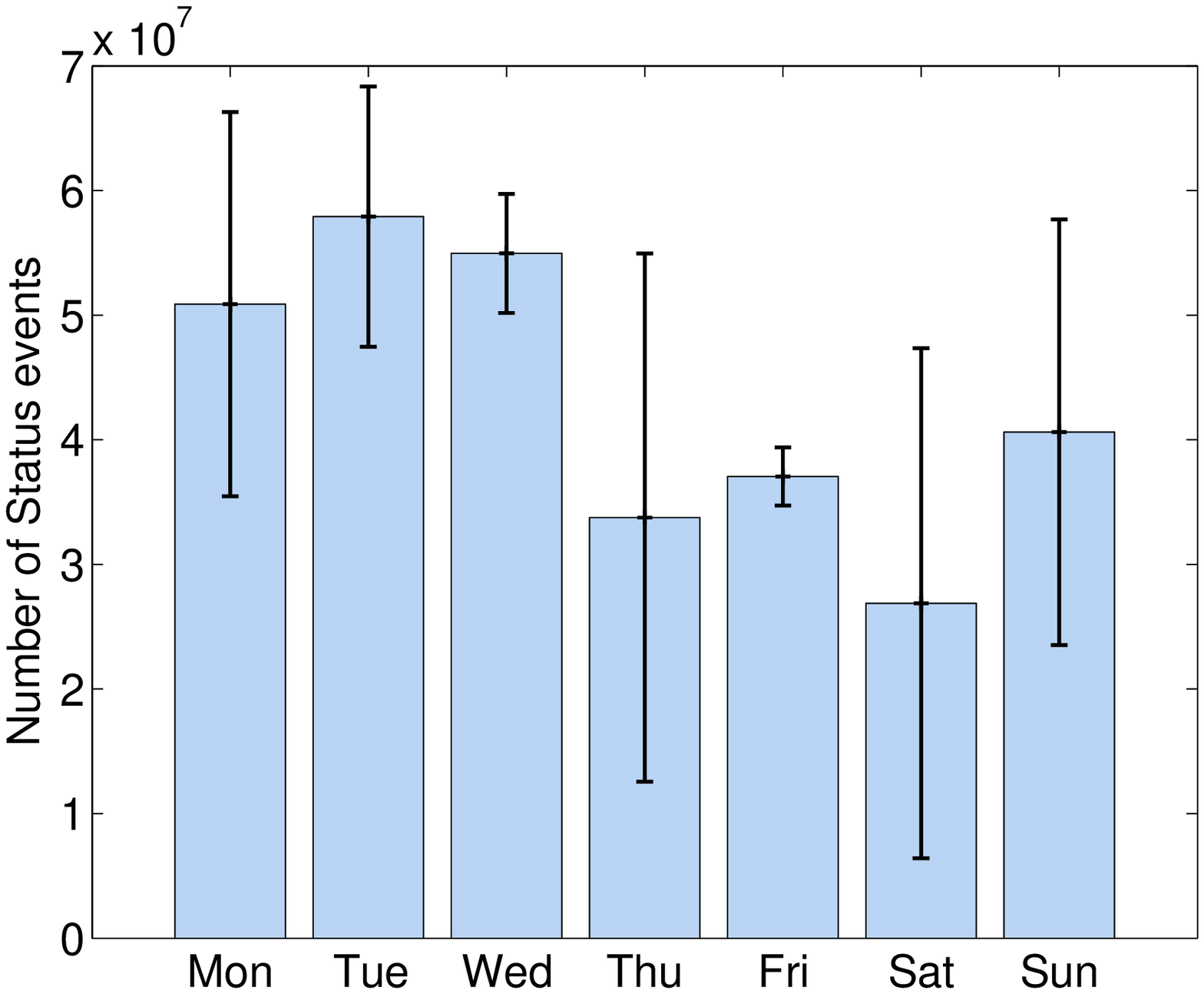} &
    \includegraphics[width=0.32\textwidth]{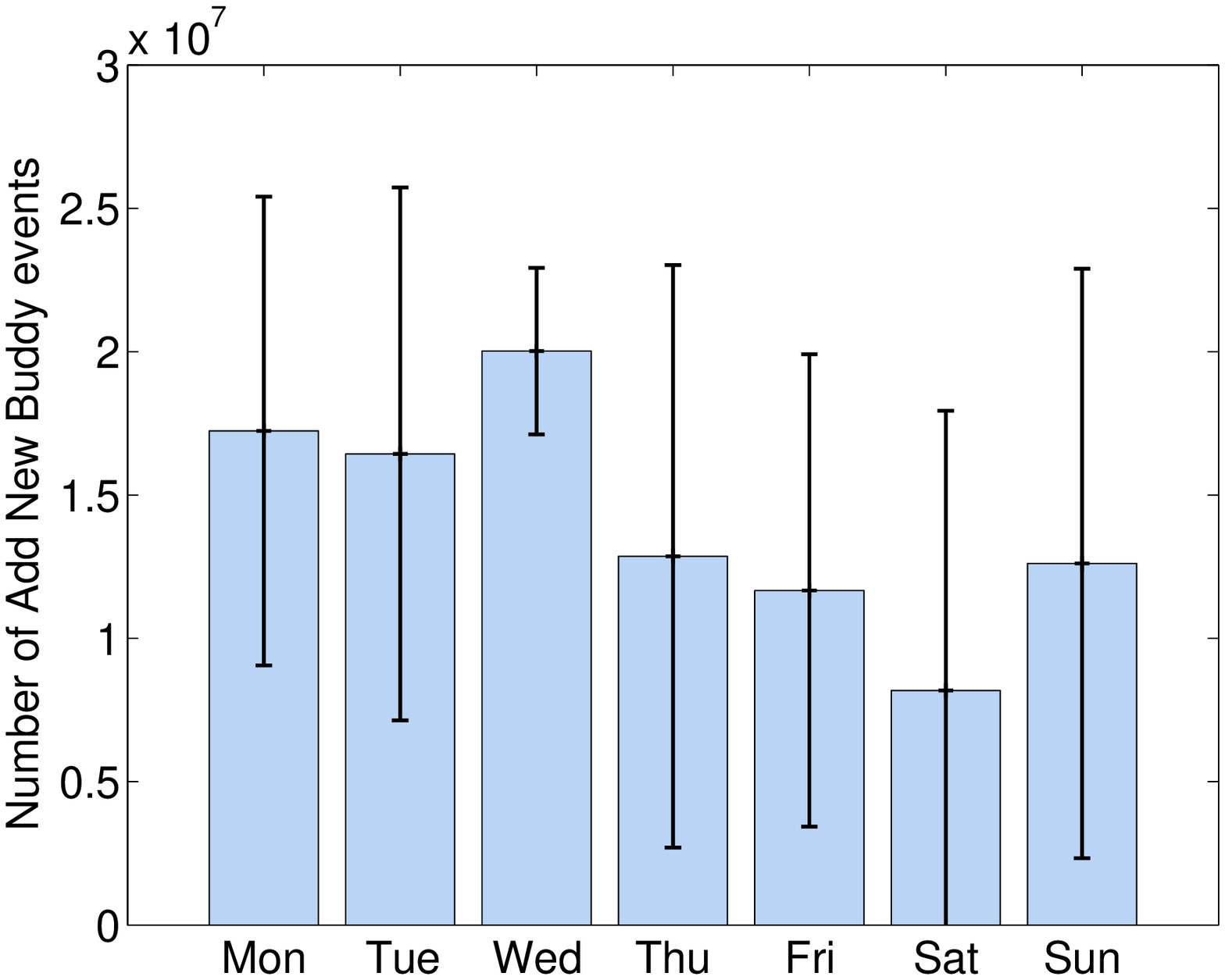} \\
    (a) Login & (b) Status & (c) Add buddy \\
  \end{tabular}
  \caption{Number of events per day of the week. We collected the
  data over a period of 5 weeks starting on May 29 2006.}
  \label{fig:eventsOverTm}
\end{center}
\end{figure}

Weekly dynamics of MSN Messenger is also quite interesting.
Figure~\ref{fig:eventsOverTm} shows the number of logins, status change and
add buddy events by day of the week over a period of 5 weeks starting in June
2006. We count the number of particular events per day of the week, and we
use the data from 5 weeks to compute the error bars.
Figure~\ref{fig:eventsOverTm}(a) shows the average number of logins per day
of the week over a 5 week period. Note that number of login events is larger
than the number of distinct users logging in, since a user can login multiple
times a day. Figure~\ref{fig:eventsOverTm}(b) plots the average number of
status change evens per day of the week. Status events include a group of 8
events describing the current status of the users, {\em i.e.,} away, be right
back, online, busy, idle, at lunch, and on the phone. Last,
Figure~\ref{fig:eventsOverTm}(c) shows the average number of add buddy events
per day of the week. Add buddy event is triggered every time user adds a new
contact to their contact list.

\subsection{Demographic characteristics of the users}

\begin{figure}
\begin{center}
  \includegraphics[width=0.9\textwidth]{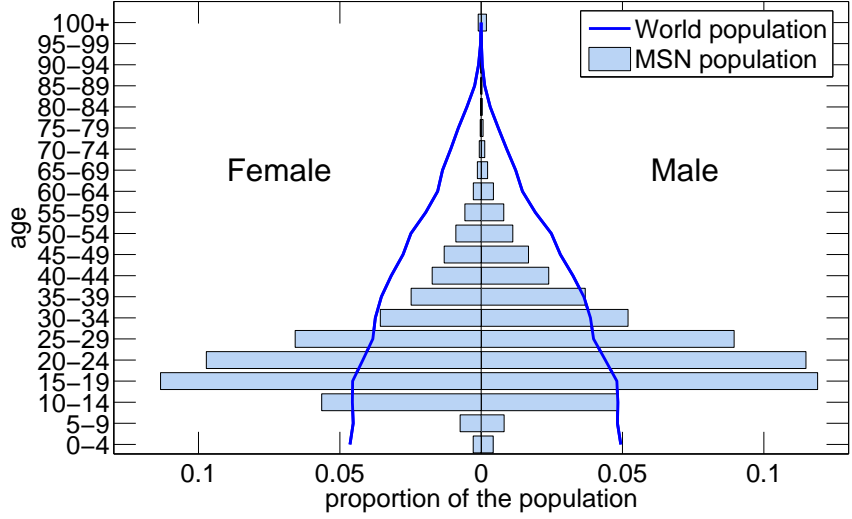}
  \caption{World and Messenger user population age pyramid.
  Ages 15--30 are overrepresented in the Messenger population.}
   \label{fig:agePyramid}
\end{center}
\end{figure}

We compared the demographic characteristics of the Messenger population with
2005 world census data and found differences between the statistics for age
and gender.  The visualization of this comparison displayed in
Figure~\ref{fig:agePyramid} shows that users with reported ages in the 15--35
span of years are strongly overrepresented in the active Messenger
population. Focusing on the differences by gender, females are
overrepresented for the 10--14 age interval. For male users, we see overall
matches with the world population for age spans 10--14 and 35—39; for women
users, we see a match for ages in the span of 30--34. We note that 6.5\% of
the population did not submit an age when creating their Messenger accounts.

\begin{figure}
\begin{center}
  \begin{tabular}{cc}
    \includegraphics[width=0.42\textwidth]{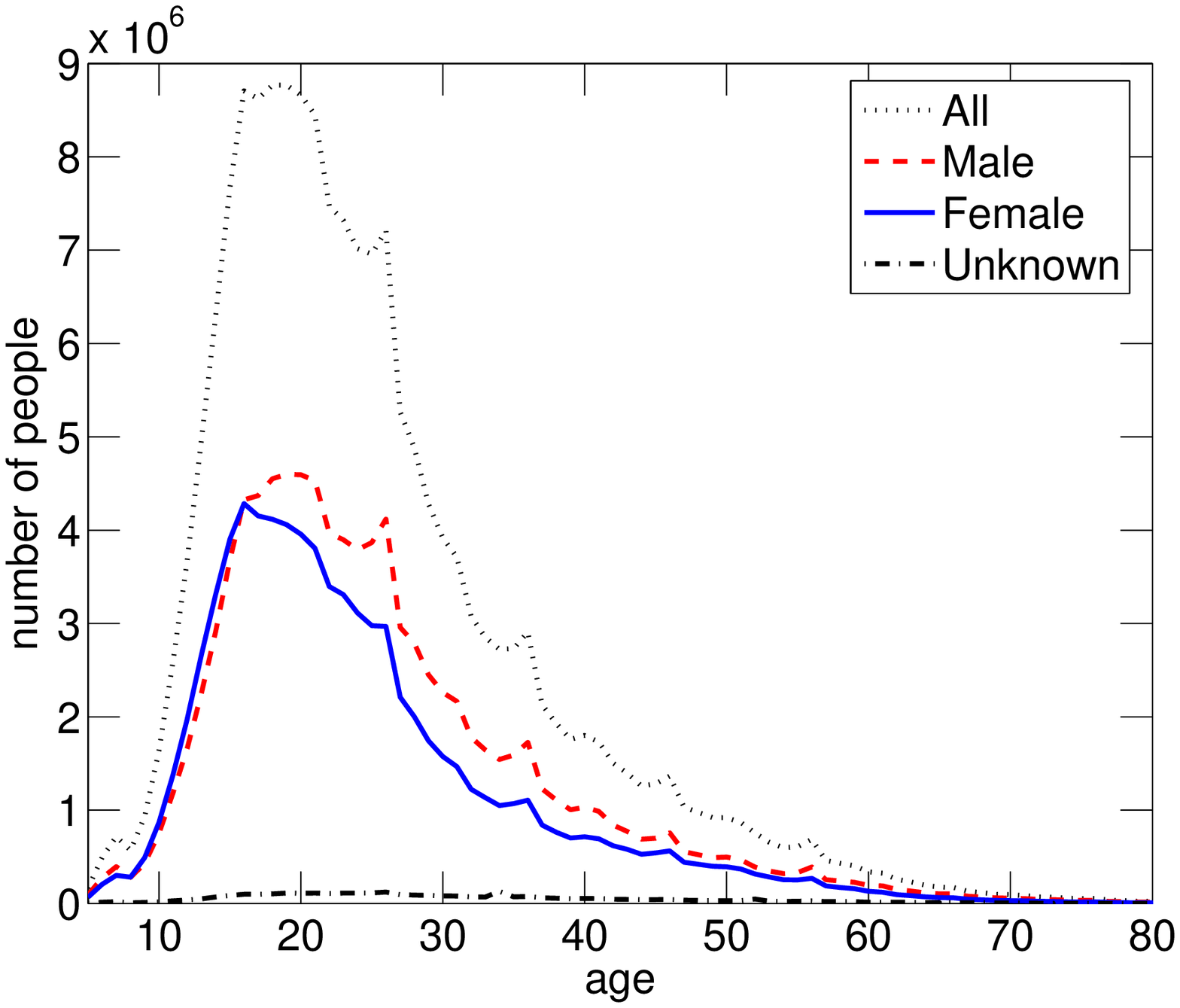} &
    \includegraphics[width=0.45\textwidth]{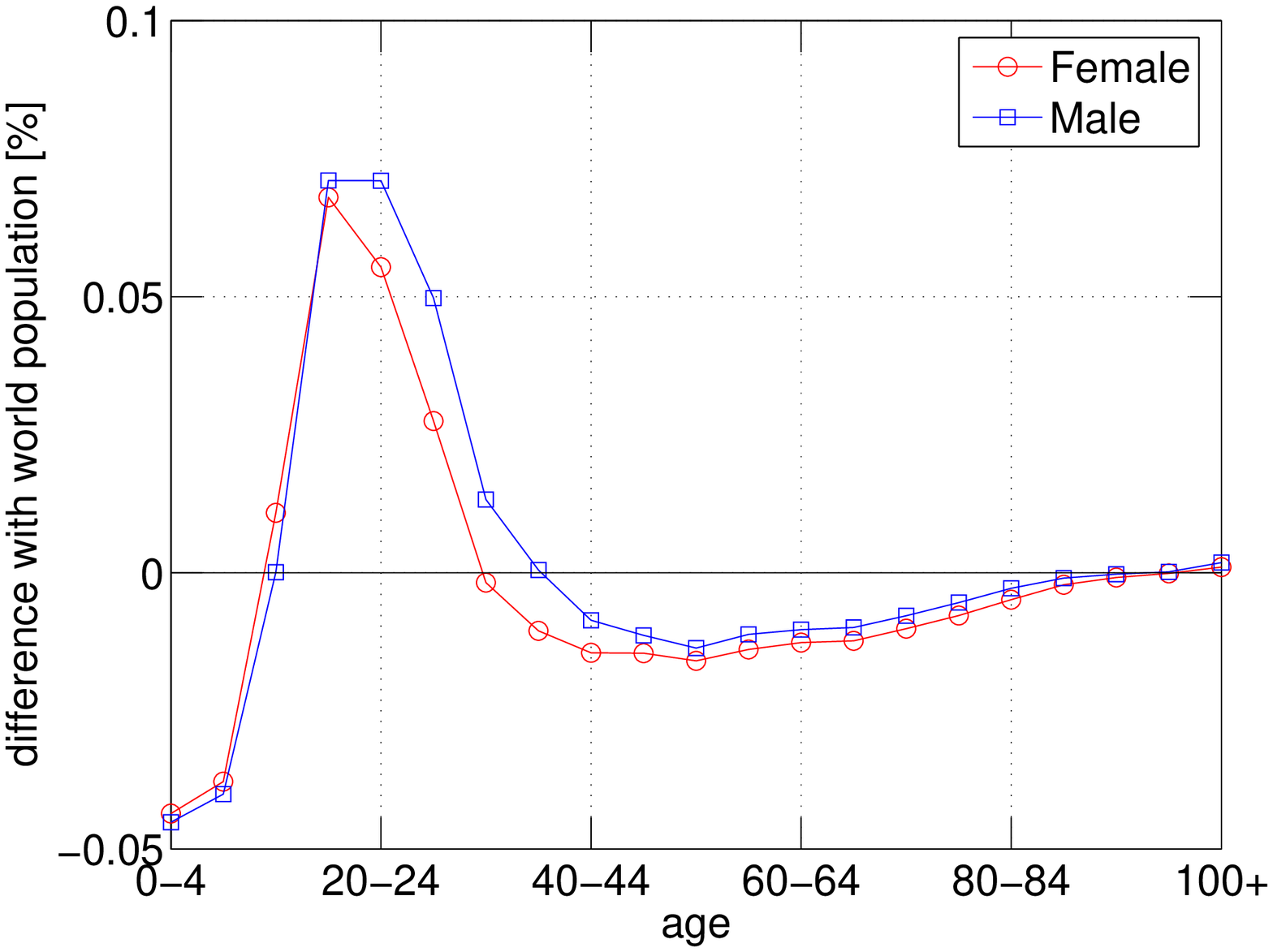} \\
         (a) Age distribution & (b) Age difference \\
  \end{tabular}
  \caption{Distribution of self-reported ages of Messenger users and the
  difference of ages of Messenger population with the world population.
  (a) Age distribution for all users, females, males and unknown users.
  (b) Relative difference of Messenger population and the world
  population. Ages 15--30 are over-represented in the Messenger
  user population.}
  \label{fig:userAge}
\end{center}
\end{figure}

To further illustrate the points above Figure~\ref{fig:userAge} shows
self-reported user age distribution and the percent difference of particular
age-group between MSN and the world population. The distribution is skewed to
the right and has a mode at age of 18. We also note that the distribution has
exponential tails.

\section{Communication characteristics}

We now focus on characteristics and patterns with communications. We limit
the analysis to conversations between two participants, which account for
99\% of all conversations.

\begin{figure}[t]
\begin{center}
  \begin{tabular}{cc}
    \includegraphics[width=0.45\textwidth]{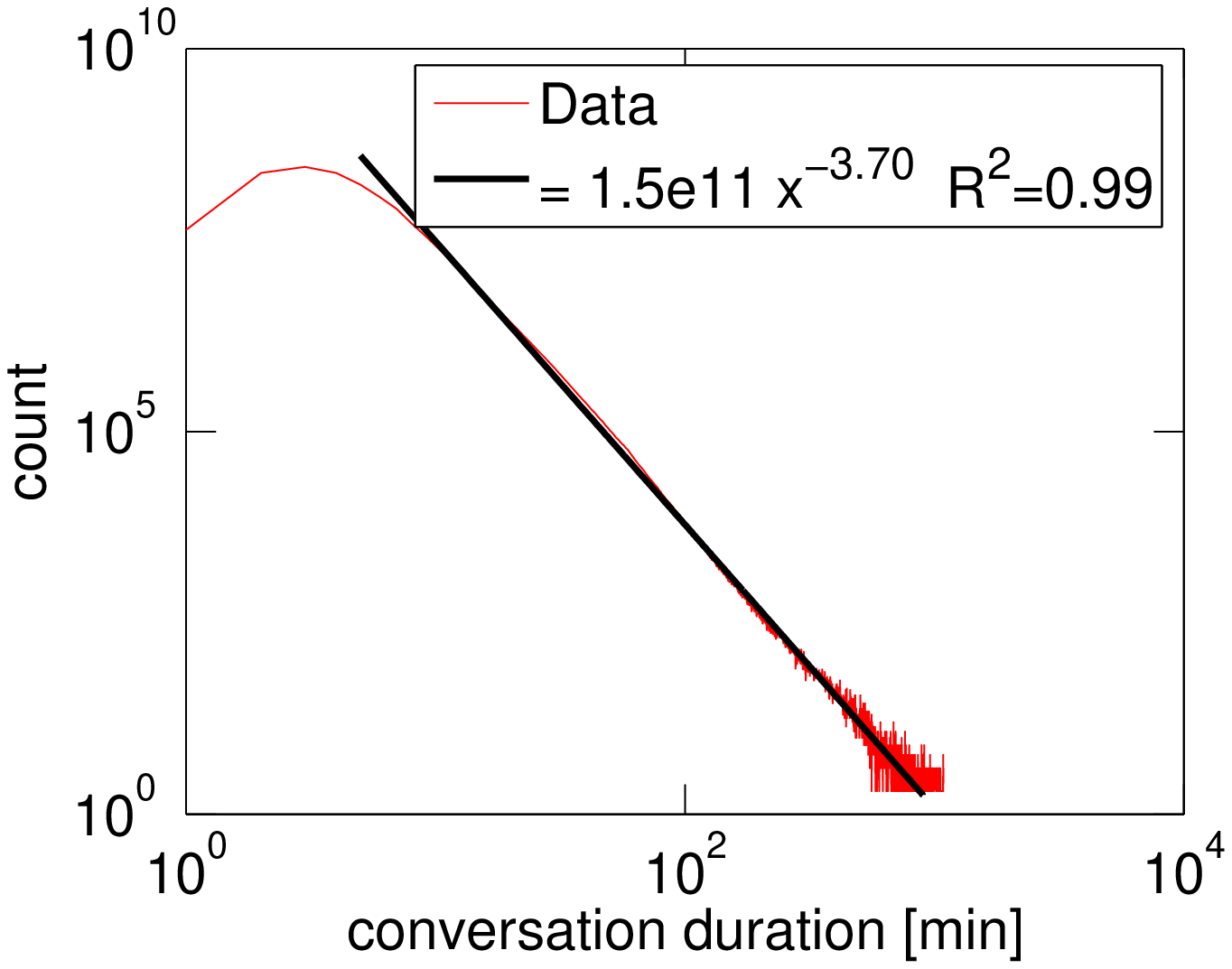} &
    \includegraphics[width=0.45\textwidth]{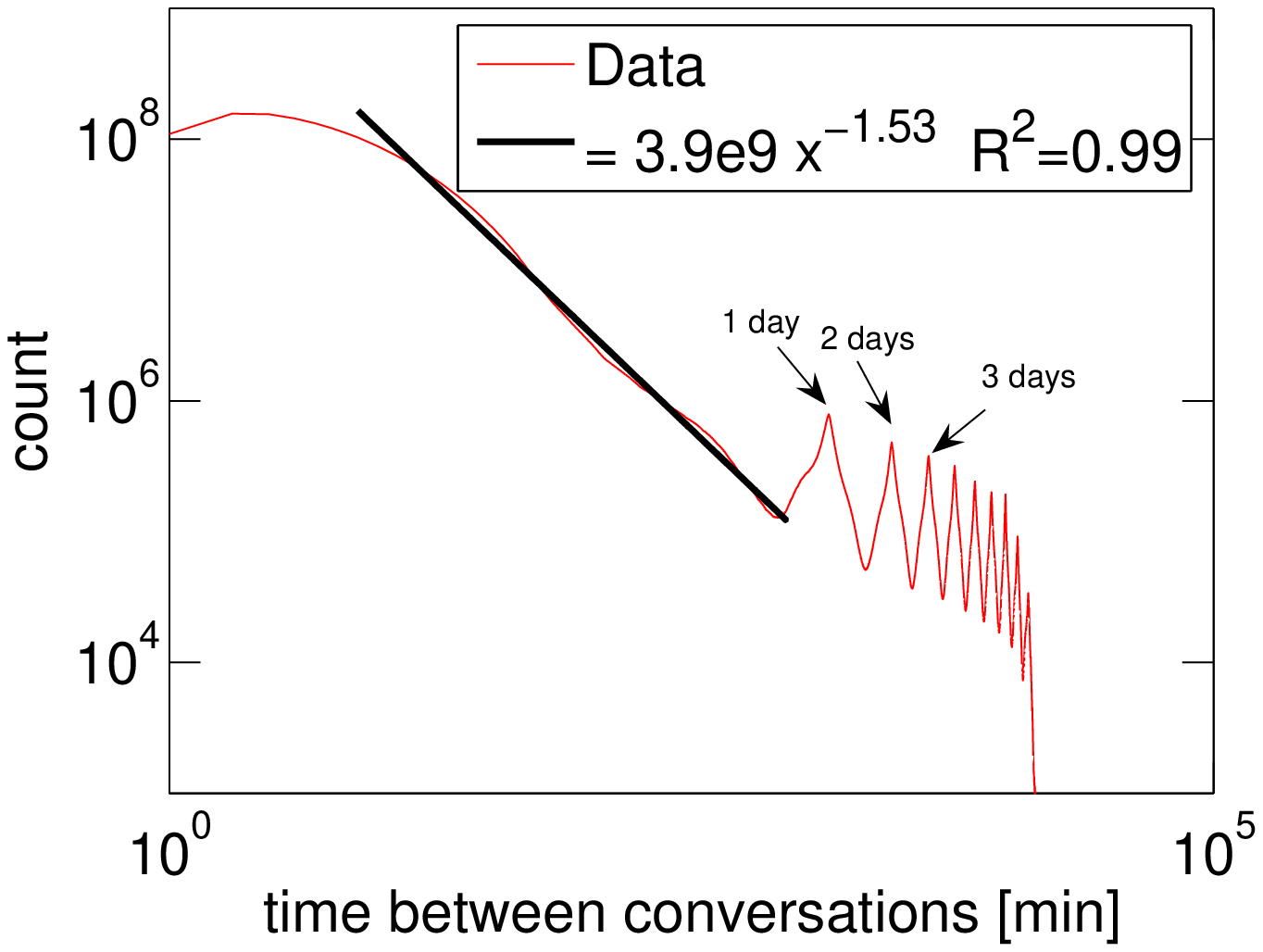} \\
  \end{tabular}
  \caption{Temporal characteristics of conversations. (a) Average
  conversation duration per user; (b) time between conversations of users.}
  \label{fig:commAgg}
\end{center}
\end{figure}

We first examine the distributions over conversation durations and times
between conversations. Let user $u$ have $C$ conversations in the observation
period. Then, for every conversation $i$ of user $u$ we create a tuple
$(ts_{u,i}, te_{u,i}, m_{u,i})$, where $ts_{u,i}$ denotes the start time of
the conversation, $te_{u,i}$ is the end time of the conversation, and
$m_{u,i}$ is the number of exchanged messages between the two users. We order
the conversations by their start time ($ts_{u,i} < ts_{u,i+1}$). Then, for
every user $u$, we calculate the average conversation duration $\bar{d}(u) =
\frac{1}{C} \sum_i te_{u,i}-ts_{u,i}$, where the sum goes over all the $u$'s
conversations. Figure~\ref{fig:commAgg}(a) shows the distribution of
$\bar{d}(u)$ over all the users $u$. We find that the conversation length can
be described by a heavy-tailed distribution with exponent -3.7  and a mode of
4 minutes.

Figure~\ref{fig:commAgg}(b) shows the intervals between consecutive
conversations of a user. We plot the distribution of $ts_{u,i+1} - ts_{u,i}$,
where $ts_{u,i+1}$ and $ts_{u,i}$ denote start times of two consecutive
conversations of user $u$. The power-law exponent of the distribution over
intervals is $~-1.5$. This result is similar to the temporal distribution for
other kinds of human communication activities, {\em e.g.}, waiting times of
emails and letters before a reply is generated~\cite{barabasi05human}. The
exponent can be explained by a priority-queue model where tasks of different
priorities arrive and wait until all tasks with higher priority are
addressed. This model generates a task waiting time distribution described by
a power-law with exponent $-1.5$.

\begin{figure}
\begin{center}
  \begin{tabular}{cc}
    \includegraphics[width=0.45\textwidth]{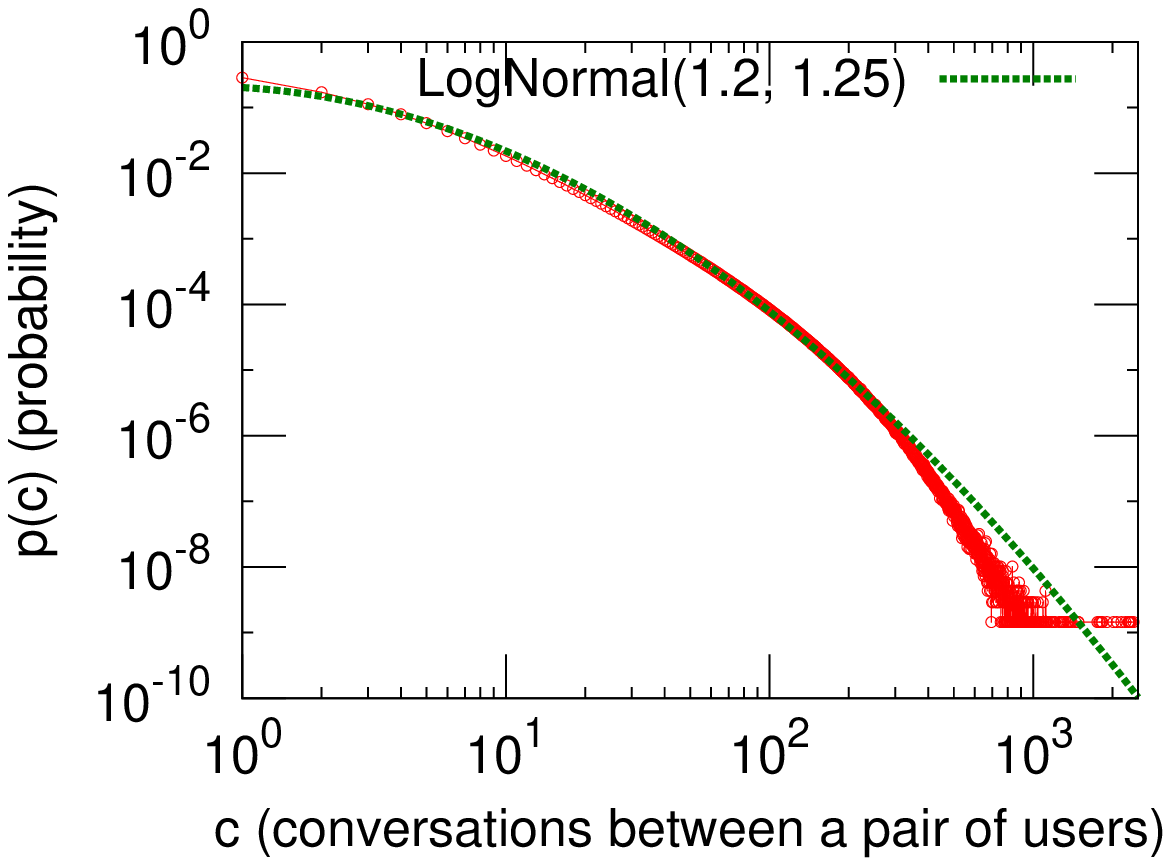} &
    \includegraphics[width=0.45\textwidth]{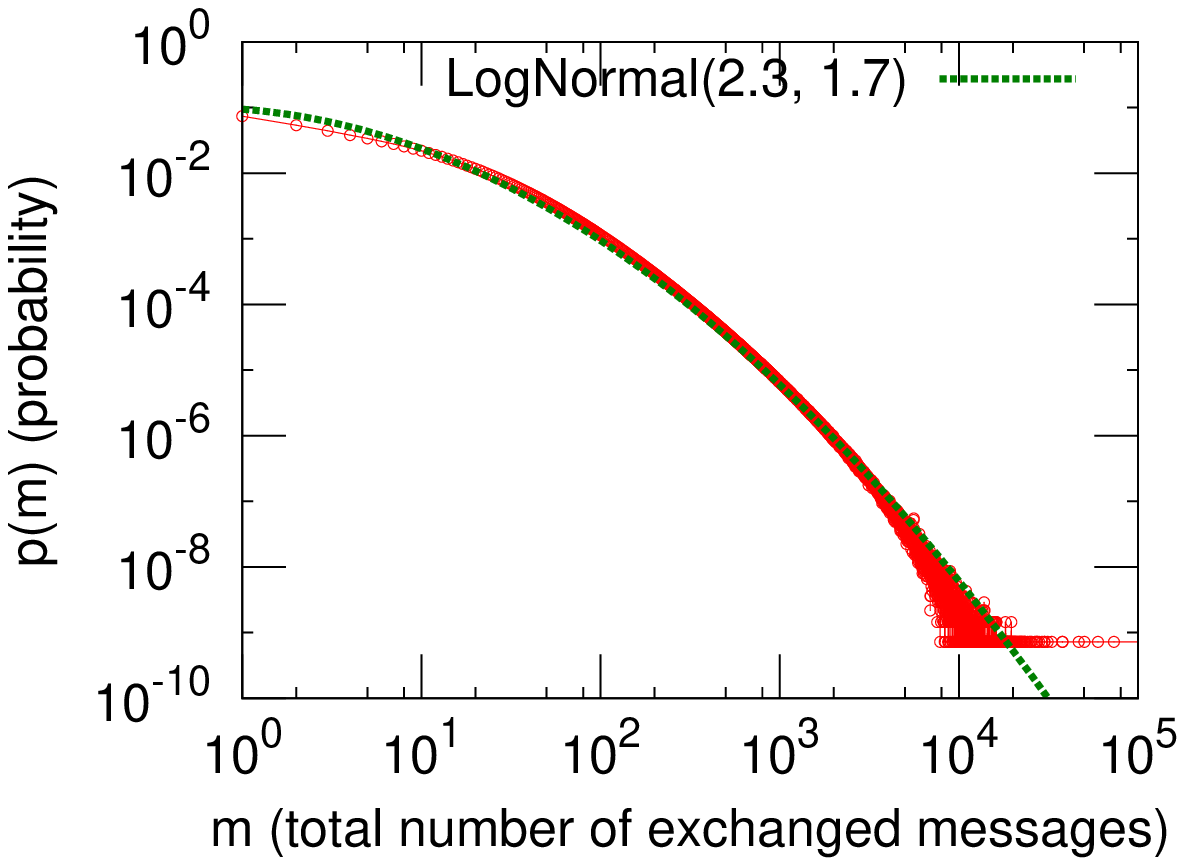} \\
     (a) Number of conversations & (b) Exchanged messages\\
  \end{tabular}
\end{center}
  \caption{Conversation statistics: (a) Number of conversations
  of a user in a month; (b) Number of messages exchanged per
  conversation;}
  \label{fig:commPairwise}
\end{figure}

However, the total number of conversations between a pair of users
(Figure~\ref{fig:commPairwise}(a)), and the total number of exchanged
messages between a pair of users (Figure~\ref{fig:commPairwise}(b)) does not
seem to follow a power law. The distribution seems still to be heavy tailed
but not power-law. The fits represent the MLE estimates of a log-normal
distribution.

\section{Communication demographics}

Next we examine the interplay of communication and user demographic
attributes, {\em i.e.}, how geography, location, age, and gender influence
observed communication patterns.

\subsection{Communication by age}

\begin{figure}
\begin{center}
  \begin{tabular}{cc}
    \includegraphics[width=0.45\textwidth]{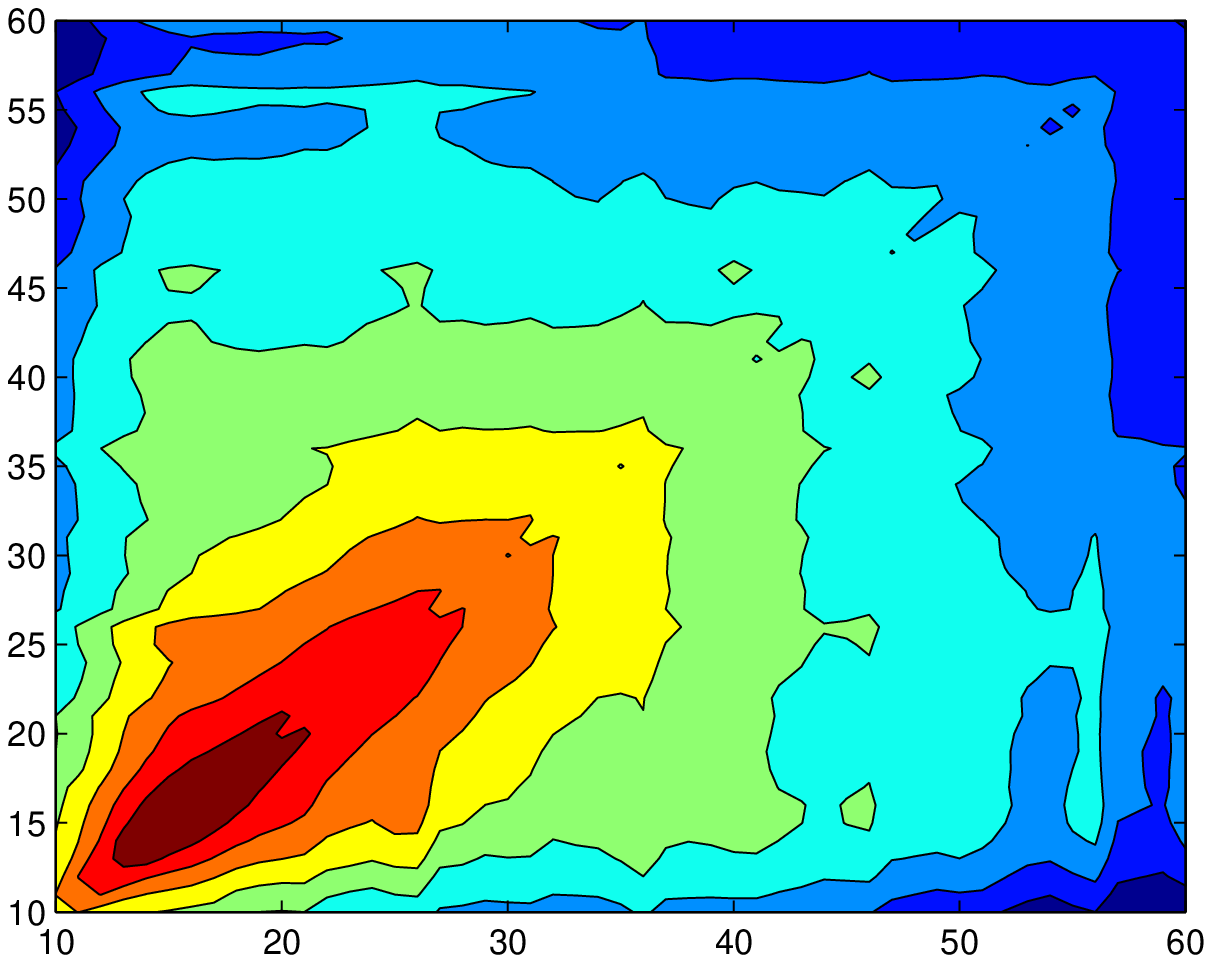} &
    \includegraphics[width=0.45\textwidth]{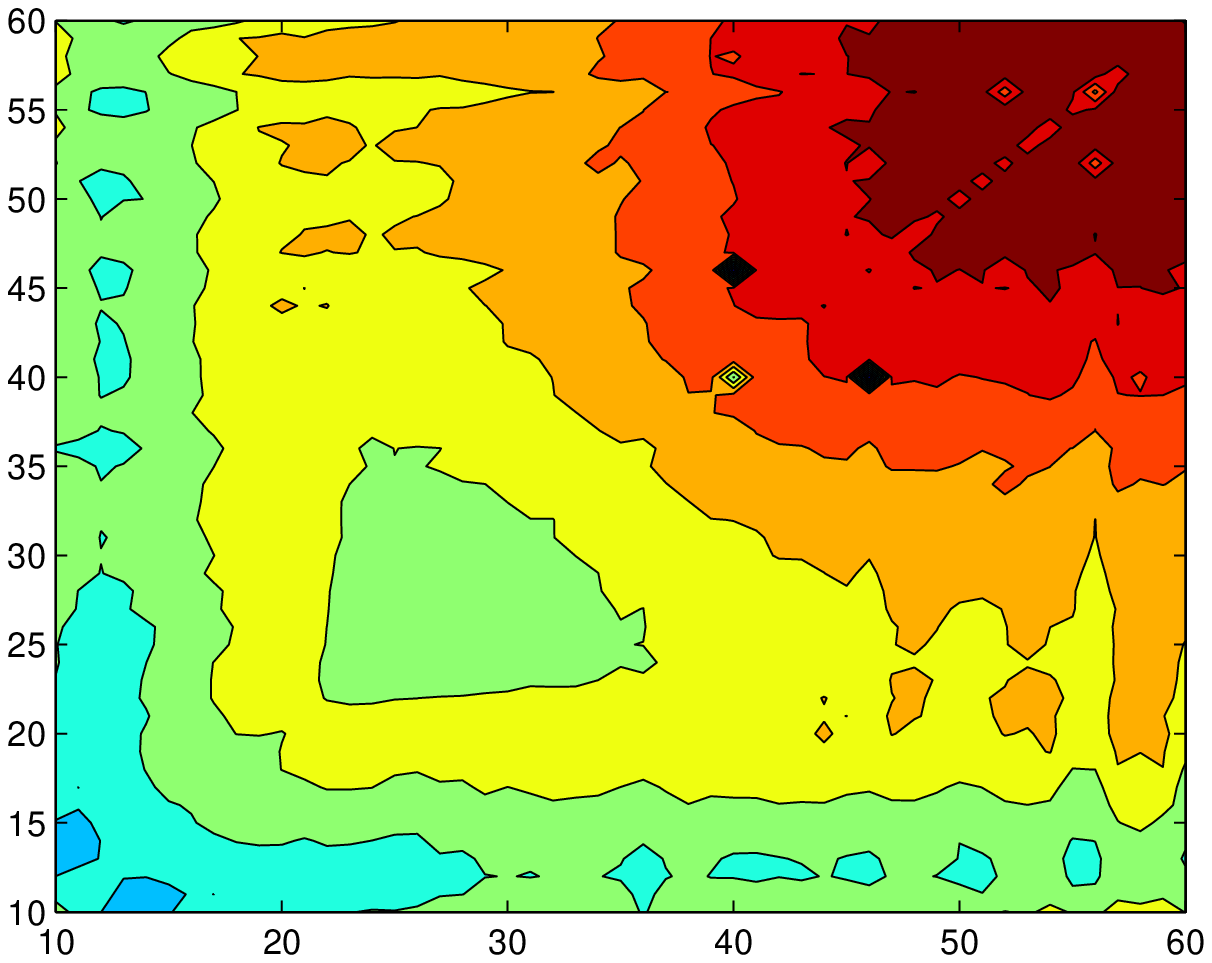} \\
    (a) Number of conversations & (b) Conversation duration\\
    \includegraphics[width=0.45\textwidth]{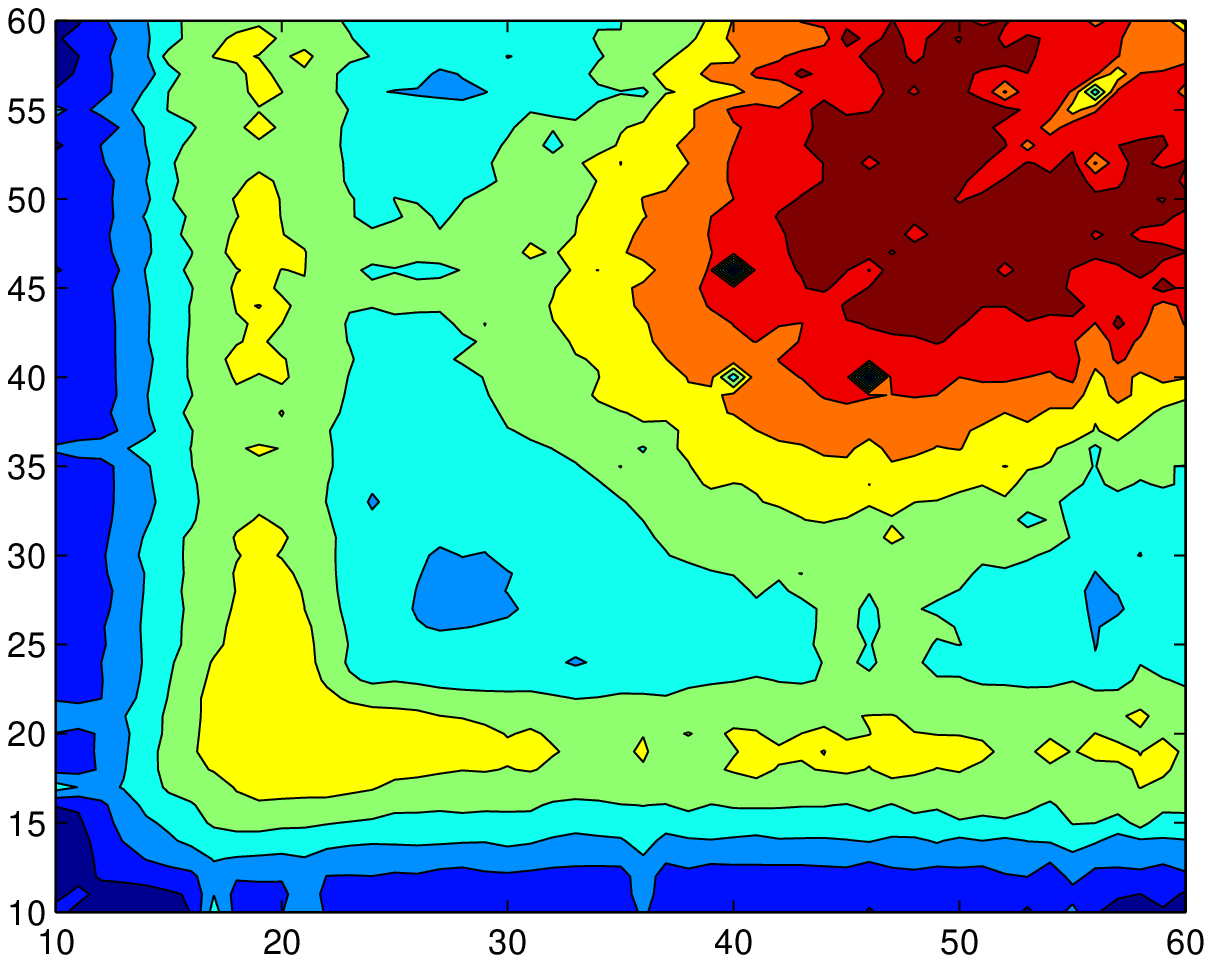} &
    \includegraphics[width=0.45\textwidth]{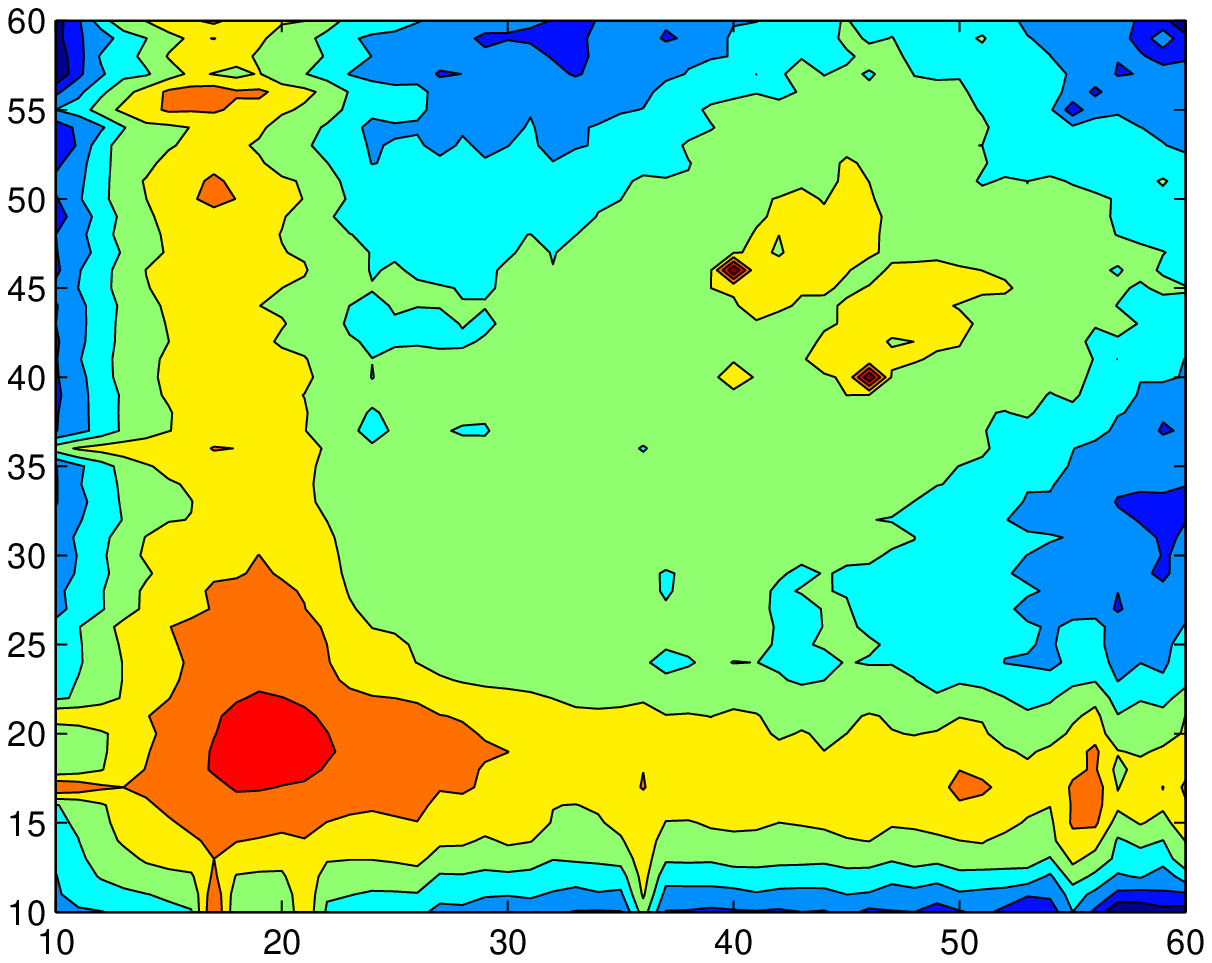} \\
    (c) Messages per conversation & (d) Messages per unit time \\
  \end{tabular}
  \caption{Communication characteristics of users by reported age. We plot
  age vs. age and the color (z-axis) represents the intensity of communication.}
  \label{fig:commAge}
\end{center}
\end{figure}

We sought to understand how communication among people changes with the
reported ages of participating users.  Figures~\ref{fig:commAge}(a)-(d) use a
heat-map visualization to communicate properties for different age--age
pairs. The rows and columns represent the ages of both parties participating,
and the color at each age--age cell captures the logarithm of the value for
the pairing. The color spectrum extends from blue (low value) through green,
yellow, and onto red (the highest value). Because of potential misreporting
at very low and high ages, we concentrate on users with self-reported ages
that fall between 10 and 60 years.

Let a tuple $(a_i, b_i, d_i, m_i)$ denote the $i$th conversation in the
entire dataset that occurred among users of ages $a_i$ and $b_i$. The
conversation had a duration of $d_i$ seconds during which $m_i$ messages were
exchanged. Let $C_{a,b} = \{(a_i, b_i, d_i, m_i) : a_i = a \wedge b_i = b\}$
denote a set of all conversations between users of ages $a$ and $b$,
respectively.

Figure~\ref{fig:commAge}(a) shows the number of conversations among people of
different ages. For every pair of ages $(a, b)$ the color indicates the size
of set $C_{a,b}$, {\em i.e.}, the number of different conversations between
users of ages $a$ and $b$. We note that, as the notion of a conversation is
symmetric, the plots are symmetric. Most conversations occur between people
of ages 10 to 20. The diagonal trend indicates that people tend to talk to
people of similar age. This is true especially for age groups between 10 and
30 years. We shall explore this observation in more detail in
Section~\ref{sec:homophily}.

Figure~\ref{fig:commAge}(b) displays a heat map for the average conversation
duration, computed as $\frac{1}{|C_{a,b}|}\sum_{i \in C_{a,b}} d_i$. We note
that older people tend to have longer conversations. We observe a similar
phenomenon when plotting the average number of exchanged messages per
conversation, computed as $\frac{1}{|C_{a,b}|}\sum_{i \in C_{a,b}} m_i$,
displayed in Figure~\ref{fig:commAge}(c). Again, we find that older people
exchange more messages, and we observe a dip for ages 25--45 and a slight
peak for ages 15--25. Figure~\ref{fig:commAge}(d) displays the number of
exchanged messages per unit time; for each age pair, $(a,b)$, we measure
$\frac{1}{|C_{a,b}|}\sum_{i \in C_{a,b}} \frac{m_i}{d_i}$. Here, we see that
younger people have faster-paced dialogs, while older people exchange
messages at a slower pace.

We note that the younger population (ages 10--35) are strongly biased towards
communicating with people of a similar age (diagonal trend in
Figure~\ref{fig:commAge}(a)), and that users who report being of ages 35
years and above tend to communicate more evenly across ages (rectangular
pattern in Fig.~\ref{fig:commAge}(a)). Moreover, older people have
conversations of the longest durations, with a ``valley'' in the duration of
conversations for users of ages 25--35. Such a dip may represent shorter,
faster-paced and more intensive conversations associated with work-related
communications, versus more extended, slower, and longer interactions
associated with social discourse.

\subsection{Communication by gender}

We report on analyses of properties of pairwise communications as a function
of the self-reported gender of users in conversations in
Table~\ref{tab:commGendLen}. Let $C_{g,h} = \{(g_i, h_i, d_i, m_i) : g_i=g
\wedge h_i=h \}$ denote a set of conversations where the two participating
users are of genders $g$ and $h$. Note that $g$ takes 3 possible values:
female, male, and unknown (unreported).

\begin{table}
  \begin{center}
  \begin{tabular}{cc}
  \begin{tabular}{c|c|c|c}
          & Unknown & Female& Male\\ \hline
      Unknown & 1.3 & 3.6 & 3.7 \\ \hline
      Female & & 21.3 & 49.9  \\ \hline
      Male   &  &   & 20.2 \\
  \end{tabular}
  &
  \begin{tabular}{c|c|c|c}
  & Unknown & Female& Male\\ \hline
  Unknown & 277 & 301 & 277 \\ \hline
  Female & & 275 & 304  \\ \hline
  Male   &  &   & 252 \\
  \end{tabular} \vspace{3mm} \\
   (a) Conversations & (b) Conversation duration \\
   & \\
  \begin{tabular}{c|c|c|c}
        & Unknown & Female& Male\\ \hline
      Unknown & 5.7 & 7.1 & 6.7 \\ \hline
      Female & & 6.6 & 7.6  \\ \hline
      Male   &  &   & 5.9 \\
  \end{tabular}
  &
  \begin{tabular}{c|c|c|c}
      & Unknown & Female& Male\\ \hline
      Unknown & 1.25 & 1.42 & 1.38 \\ \hline
      Female & & 1.43 & 1.50 \\ \hline
      Male   &  &  & 1.42 \\
  \end{tabular} \vspace{3mm} \\
   (c) Exchanged messages per conversation & (d) Conversation intensity\\
  \end{tabular}
  \end{center}
  \caption{Cross-gender communication. Data is based on all
  two-person conversations from June 2006. (a) Percentage of conversations
  among users of different self-reported gender; (b) average
  conversation length in seconds; (c) number of exchanged messages per
  conversation; (d) number of exchanged messages per minute of
  conversation.}
  \label{tab:commGendLen}
\end{table}

Table~\ref{tab:commGendLen}(a) relays $|C_{g,h}|$ for combinations of genders
$g$ and $h$. The table shows that approximately 50\% of conversations occur
between male and female and 40\% of the conversations occur among users of
the same gender (20\% for each). A small number of conversations occur
between people who did not reveal their gender.

Similarly, Table~\ref{tab:commGendLen}(b) shows the average conversation
length in seconds, broken down by the gender of conversant, computed as
$\frac{1}{|C_{g,h}|} \sum_{i\in C_{g,h}} d_i$. We find that male--male
conversations tend to be shortest, lasting approximately 4 minutes.
Female--female conversations last 4.5 minutes on the average. Female--male
conversations have the longest durations, taking more than 5 minutes on
average. Beyond taking place over longer periods of time, more messages are
exchanged in female--male conversations. Table~\ref{tab:commGendLen}(c) lists
values for $\frac{1}{|C_{g,h}|} \sum_{i\in C_{g,h}} m_i$ and shows that, in
female--male conversations, 7.6 messages are exchanged per conversation on
the average as opposed to 6.6 and 5.9 for female--female and male--male,
respectively. Table~\ref{tab:commGendLen}(d) shows the communication
intensity computed as $\frac{1}{|C_{g,h}|} \sum_{i\in C_{g,h}}
\frac{m_i}{d_i}$. The number of messages exchanged per minute of conversation
for male--female conversations is higher at 1.5 messages per minute than for
cross-gender conversations, where the rate is 1.43 messages per minute.

We examined the number of {\em communication ties}, where a tie is
established between two people when they exchange at least one message during
the observation period. We computed 300 million male--male ties, 255 million
female--female ties, and 640 million cross-gender ties. The Messenger
population consists of 100 million males and 80 million females by self
report. These findings demonstrate that ties are not heavily gender biased;
based on the population, random chance predicts 31\% male--male, 20\%
female--female, and 49\% female--male links. We observe 25\% male--male, 21\%
female--female, and 54\% cross-gender links, thus demonstrating a minor bias
of female--male links.

The results reported in Table~\ref{tab:commGendLen} run counter to prior
studies reporting that communication among individuals who resemble one other
(same gender) occurs more often (see~\cite{mcpherson01homophily} and
references therein). We identified significant heterophily, where people tend
to communicate more with people of the opposite gender. However, we note that
link heterogeneity was very close to the population
value~\cite{marsden87core}, {\em i.e.}, the number of same- and cross-gender
ties roughly corresponds to random chance. This shows there is no significant
bias in linking for gender. However, we observe that cross-gender
conversations tend to be longer and to include more messages, suggesting that
more effort is devoted to conversations with the opposite sex.

\subsection{World geography and communication}

\begin{figure}[t]
\begin{center}
  \includegraphics[width=0.9\textwidth]{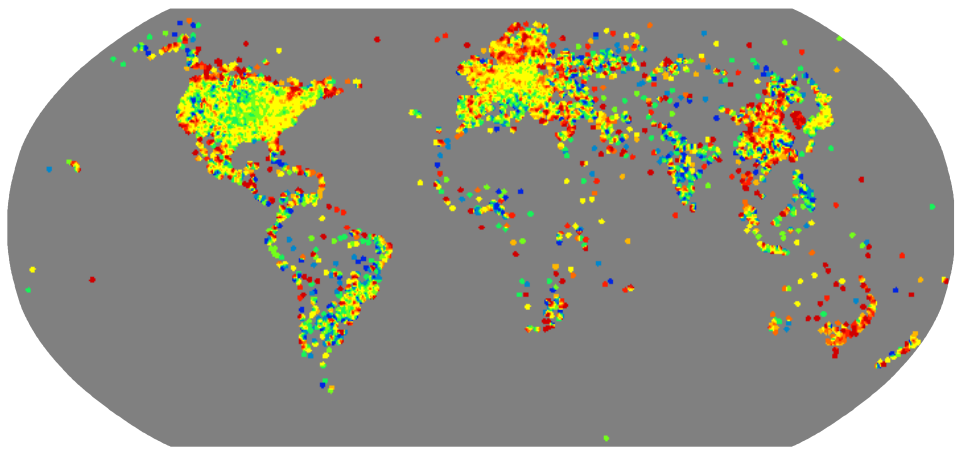}
  \caption{Number of users at a particular geographic location.
  Color represents the number of users. Notice the map of
  the world appears.}
  \label{fig:usersMap}
\end{center}
\end{figure}

We now focus on the influence of geography and distance among participants on
communications. Figure~\ref{fig:usersMap} shows the geographical locations of
Messenger users. The general location of the user was obtained via reverse IP
lookup. We plot all latitude/longitude positions linked to the position of
servers where users log into the service. The color of each dot corresponds
to the logarithm of the number of logins from the respective location, again
using a spectrum of colors ranging from blue (low) through green and yellow
to red (high). Although the maps are built solely by plotting these
positions, a recognizable world map is generated. We find that North America,
Europe, and Japan are very dense, with many users from those regions using
Messenger. For the rest of the world, the population of Messenger users
appears to reside largely in coastal regions.

\begin{figure}[t]
\begin{center}
  \includegraphics[width=0.9\textwidth]{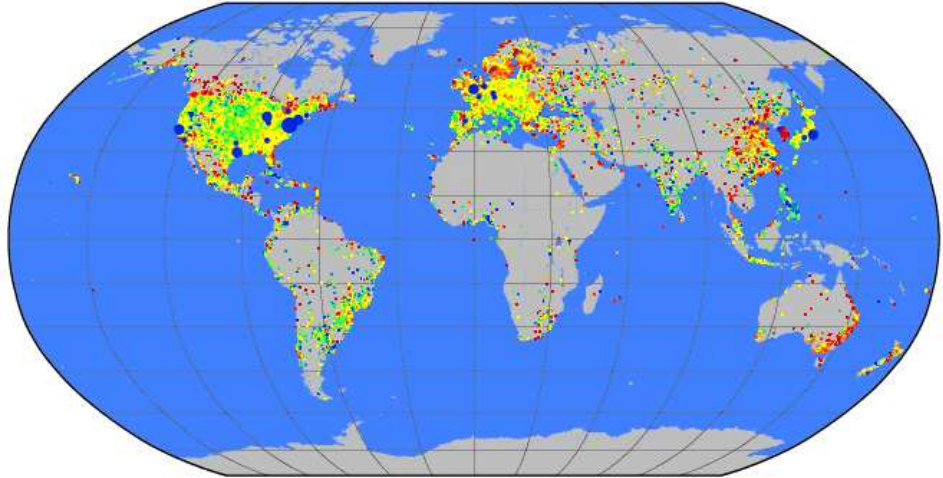}
  \caption{Number of users at particular geographic location
  superimposed on the map of the world. Color represents the number of
  users.}
  \label{fig:usersMap1}
\end{center}
\end{figure}

\begin{figure}[t]
\begin{center}
  \includegraphics[width=0.9\textwidth]{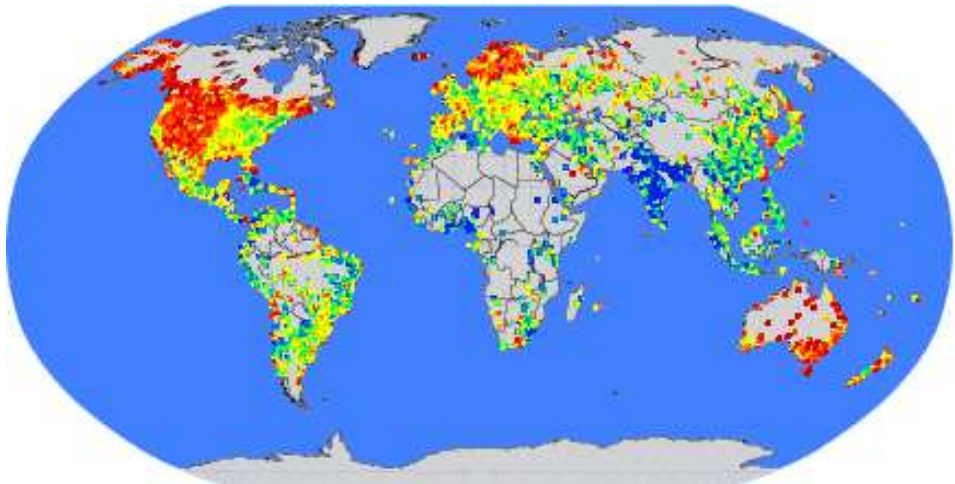}
  \caption{Number of Messenger users per capita. Color intensity
  corresponds to the number of users per capita in the cell of the grid.}
  \label{fig:usersPerCapita}
\end{center}
\end{figure}

We can condition the densities and behaviors of Messenger users on multiple
geographical and socioeconomic variables and explore relationships between
electronic communications and other attributes.  As an example, harnessed the
United Nations gridded world population data to provide estimates of the
number of people living in each cell. Given this data, and the data from
Figure~\ref{fig:usersMap}, we calculate the number of users per capita,
displayed in Figure~\ref{fig:usersPerCapita}. Now we see transformed picture
where several sparsely populated regions stand out as having a high usage per
capita.  These regions include the center of the United States, Canada,
Scandinavia, Ireland, Australia, and South Korea.

\begin{figure}[t]
\begin{center}
  \includegraphics[width=0.9\textwidth,height=6cm]{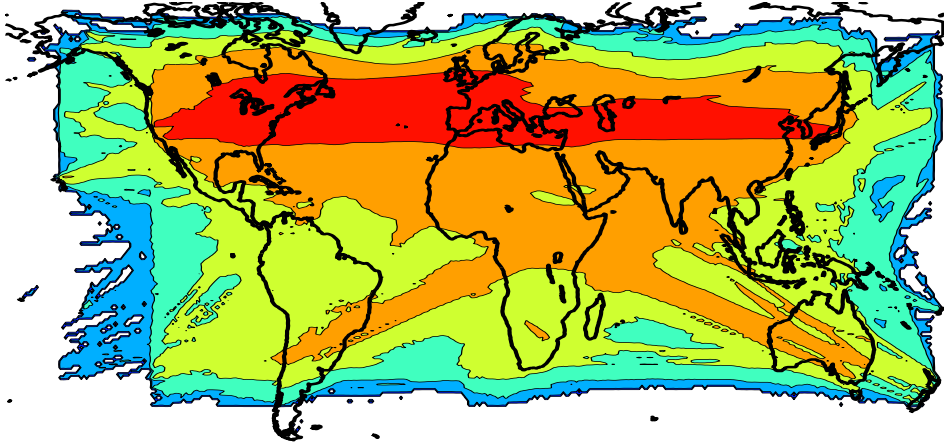}
  \caption{A communication heat map.}
  \label{fig:commHeatMap}
\end{center}
\end{figure}

Figure~\ref{fig:commHeatMap} shows a heat map that represents the intensities
of Messenger communications on an international scale. To create this map, we
place the world map on a fine grid, where each cell of the grid contains the
count of the number of conversations that pass through that point by
increasing the count of all cells on the straight line between the
geo-locations of pairs of conversants. The color indicates the number of
conversations crossing each point, providing a visualization of the key flows
of communication. For example, Australia and New Zealand have communications
flowing towards Europe and United States. Similar flows hold for Japan. We
see that Brazilian communications are weighted toward Europe and Asia. We can
also explore the flows of transatlantic and US transcontinental
communications.

\subsection{Communication among countries}

\begin{figure}
\begin{center}
  \begin{tabular}{cc}
    \includegraphics[width=0.45\textwidth]{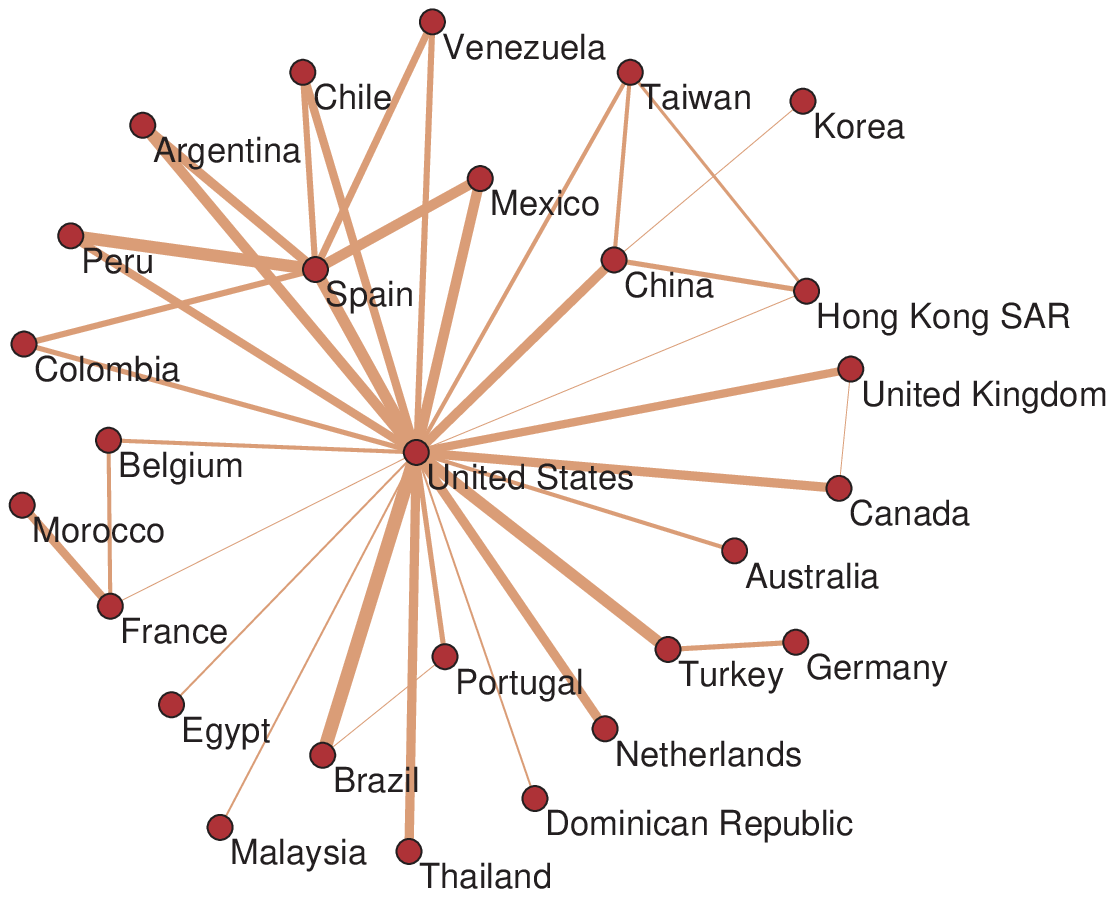} &
    \includegraphics[width=0.45\textwidth]{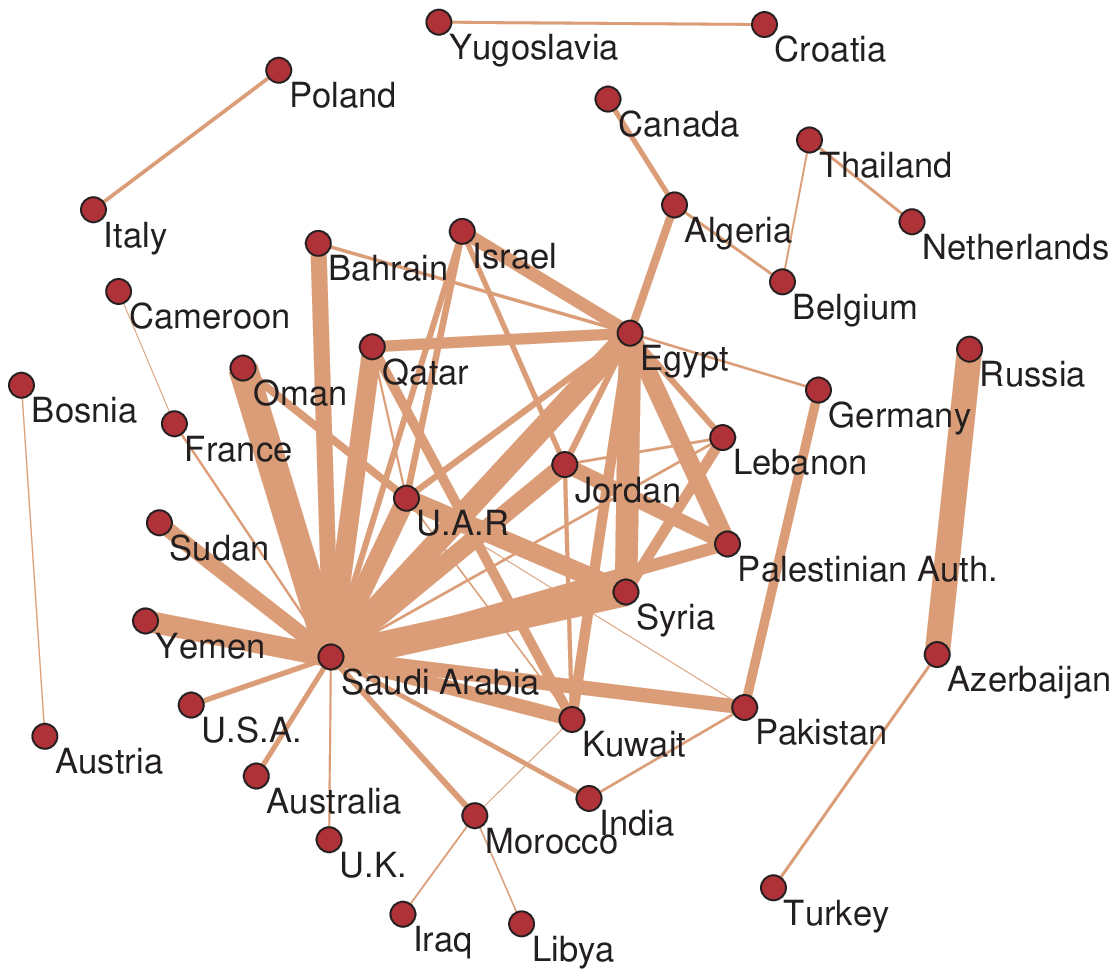} \\
  \end{tabular}
  \caption{(a) Communication among countries with at least 10
  million conversations in June 2006. (b) Countries by average length of the conversation. Edge widths correspond to logarithms of intensity of links.}
  \label{fig:cntryGraph}
\end{center}
\end{figure}

Communication among people within different countries also varies depending
on the locations of conversants. We examine two such views.
Figure~\ref{fig:cntryGraph}(a) shows the top countries by the number of
conversations between pairs of countries. We examined all pairs of countries
with more than 10 million conversations per month. The width of edges in the
figure is proportional to the logarithm of the number of conversations among
the countries. We find that the United States and Spain appear to serve as
hubs and that edges appear largely between historically or ethnically
connected countries.  As examples, Spain is connected with the Spanish
speaking countries in South America, Germany links to Turkey, Portugal to
Brazil, and China to Korea.

Figure~\ref{fig:cntryGraph}(b) displays a similar plot where we consider
country pairs by the average duration of conversations. The width of the
edges are proportional to the mean length of conversations between the
countries. The core of the network appears to be Arabic countries, including
Saudi Arabia, Egypt, United Arab Emirates, Jordan, and Syria.

\begin{table}
  \begin{center}
  \begin{tabular}{c|c}
    Country & Fraction of population\\ \hline \hline
    Iceland & 0.35   \\
    Spain & 0.28   \\
    Netherlands & 0.27  \\
    Canada & 0.26   \\
    Sweden & 0.25   \\
    Norway & 0.25   \\
    Bahamas, The & 0.24   \\
    Netherlands Antilles & 0.24  \\
    Belgium & 0.23  \\ \hline
    France & 0.18 \\
    United Kingdom & 0.17 \\
    Brazil & 0.08 \\
    United States & 0.08 \\
  \end{tabular}
  \end{center}
  \caption{Top 10 countries with most the largest number of Messenger
  users. Fraction of country's population actively using Messenger.}
  \label{tab:cntryUsers}
\end{table}

Comparing the number of active users with the country population reveals
interesting findings. Table~\ref{tab:cntryUsers} shows the top 10 countries
with the highest fraction of population using Messenger. These are mainly
northern European countries and Canada. Countries with most of the users (US,
Brazil) tend to have smaller fraction of population using Messenger.

\begin{table}
  \begin{center}
  \begin{tabular}{c|c}
  Country & Conversations per user per day \\ \hline \hline
  Afghanistan & 4.37   \\
  Netherlands Antilles & 3.79   \\
  Jamaica & 2.63   \\
  Cyprus & 2.33   \\
  Hong Kong & 2.27  \\
  Tunisia & 2.25 \\
  Serbia & 2.15 \\
  Dominican Republic & 2.06 \\
  Bulgaria & 2.07 \\
  \end{tabular}
  \end{center}
  \caption{Top 10 countries by the number of conversations per user per
  day.}
  \label{tab:cntryConvs}
\end{table}

Similarly, Table~\ref{tab:cntryConvs} shows the top 10 countries by the
number of conversations per user per day. Here the countries are very diverse
with Afghanistan topping the list. The Netherlands Antilles appears on top 10
list for both the fraction of the population using Messenger and the number
of conversations per user.

\begin{table}
  \begin{center}
  \begin{tabular}{c|c|c}
  Country & Messages per user per day & Minutes talking per user per day
  \\
  \hline \hline
  Afghanistan & 32.00 & 20.91 \\
  Netherlands Antilles & 24.12 & 17.43 \\
  Serbia & 22.41 & 12.01 \\
  Bosnia and Herzegovina & 22.40 & 11.41 \\
  Macedonia & 19.52 & 10.46 \\
  Cyprus & 19.33 & 12.37 \\
  Tunisia & 19.17 & 13.54 \\
  Bulgaria & 18.94 & 11.38 \\
  Croatia & 17.78 & 10.05 \\
  \end{tabular}
  \end{center}
  \caption{Top 10 countries by the number of messages and minutes
  talking per user per day.}
  \label{tab:cntryCommTm}
\end{table}

Last, Table~\ref{tab:cntryCommTm} shows the top 10 countries by the number of
messages and minutes talking per user per day. We note that the list of the
countries is similar to those in Table~\ref{tab:cntryConvs}. Afghanistan
still tops the list but now most of the talkative counties come from Eastern
Europe (Serbia, Bosnia, Bulgaria, Croatia).

\subsection{Communication and geographical distance}

We were interested in how communications change as the distance between
people increases. We had hypothesized that the number of conversations would
decrease with geographical distance as users might be doing less coordination
with one another on a daily basis, and where communication would likely
require more effort to coordinate than might typically be needed for people
situated more locally.  We also conjectured that, once initiated,
conversations among people who are farther apart would be somewhat longer as
there might be a stronger need to catch up when the less-frequent
conversations occurred.

\begin{figure}
  \begin{center}
  \begin{tabular}{cc}
  \includegraphics[width=0.45\textwidth]{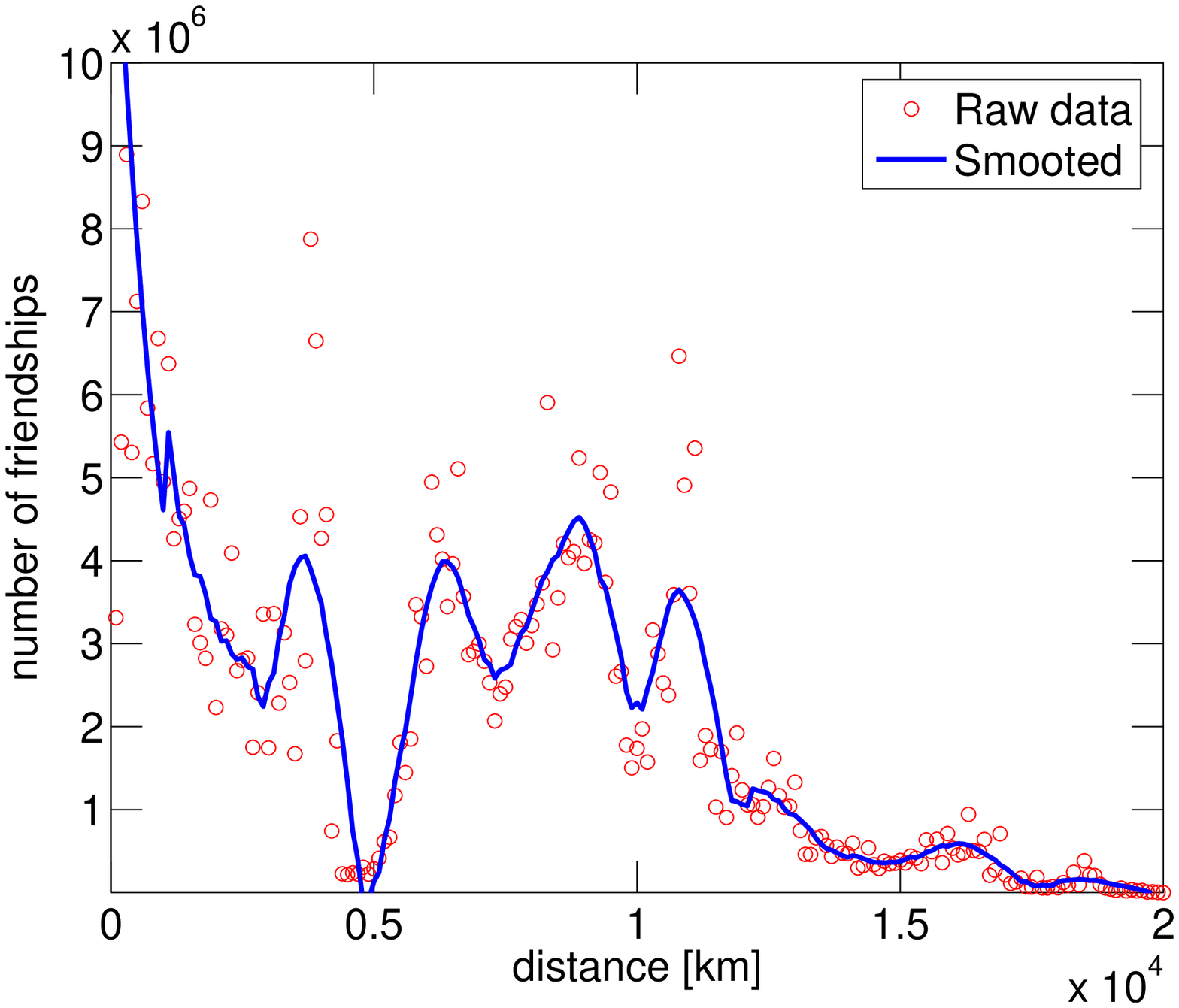} &
  \includegraphics[width=0.45\textwidth]{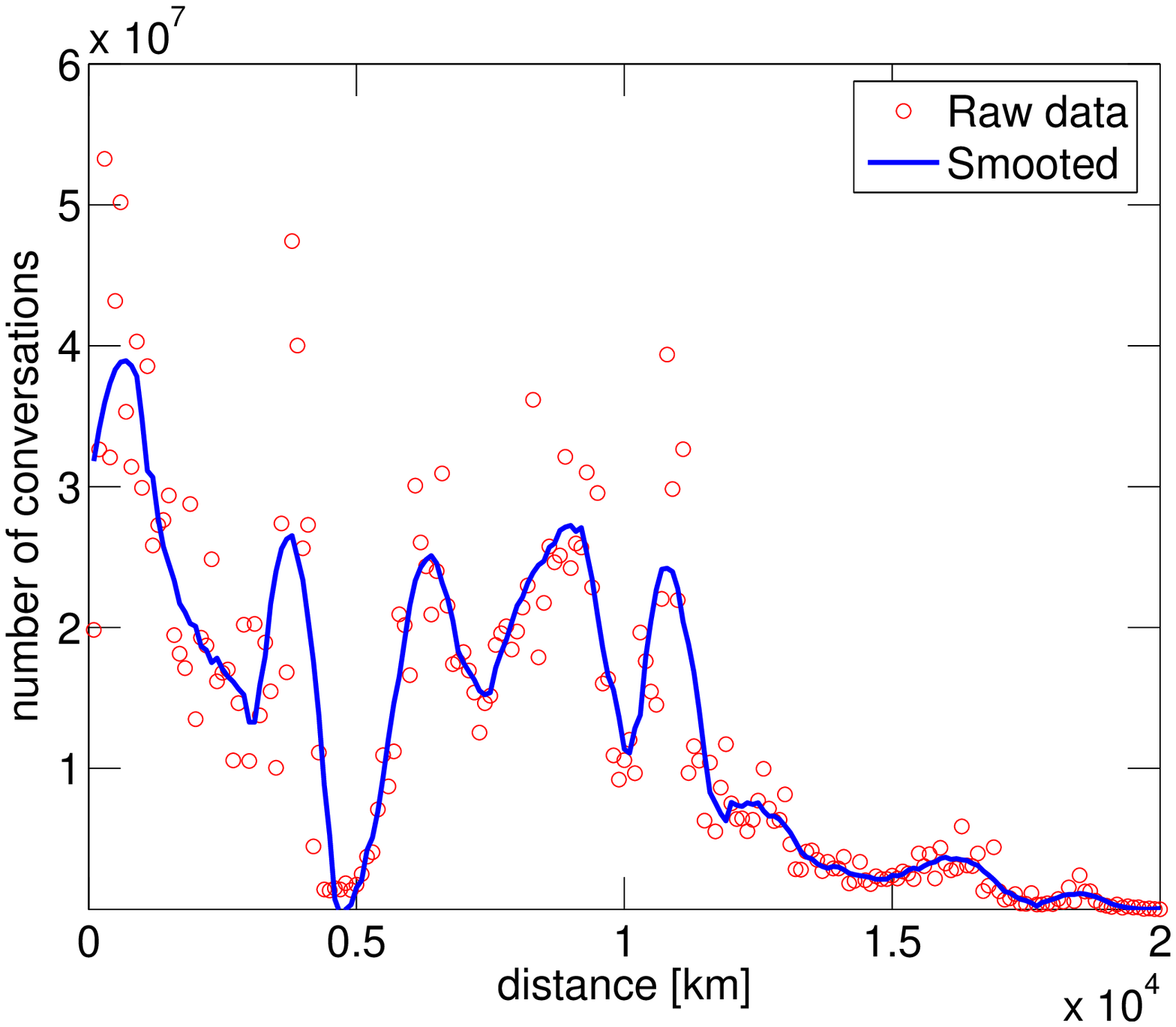} \\
  (a) Number of links & (b) Number of conversations \\
  \includegraphics[width=0.45\textwidth]{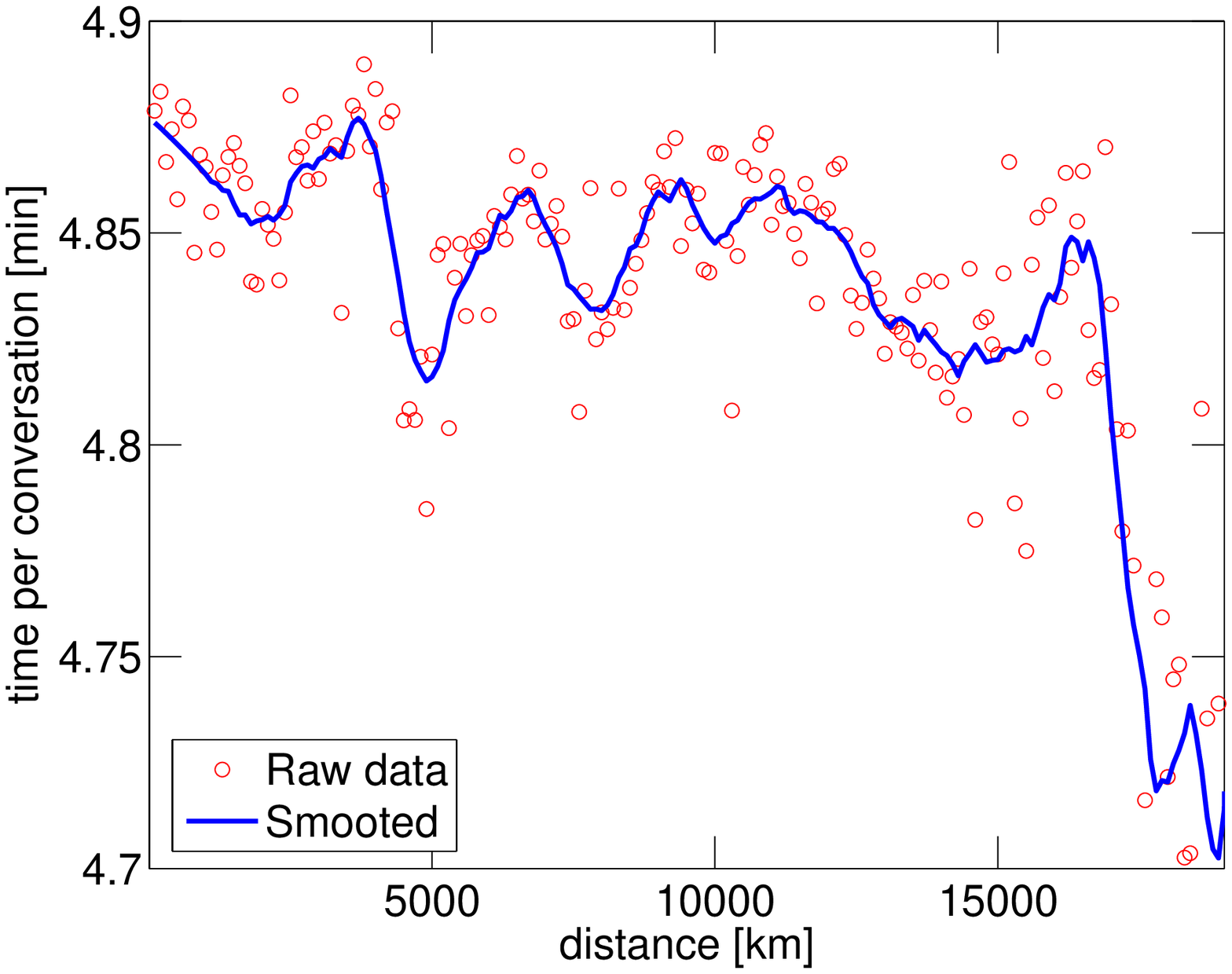} &
  \includegraphics[width=0.45\textwidth]{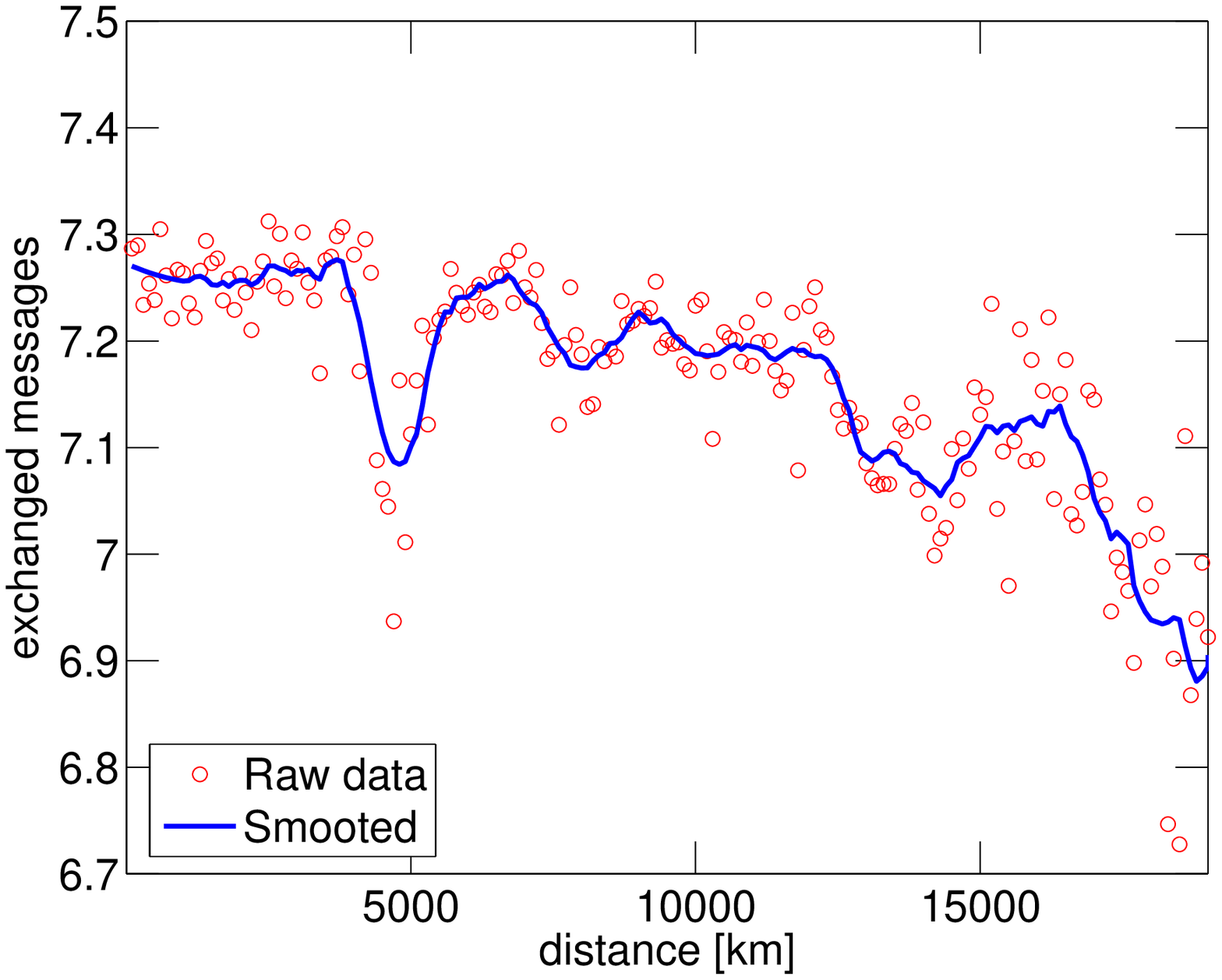} \\
  (c) Conversation duration  & (d) Exchanged messages\\
  \includegraphics[width=0.45\textwidth]{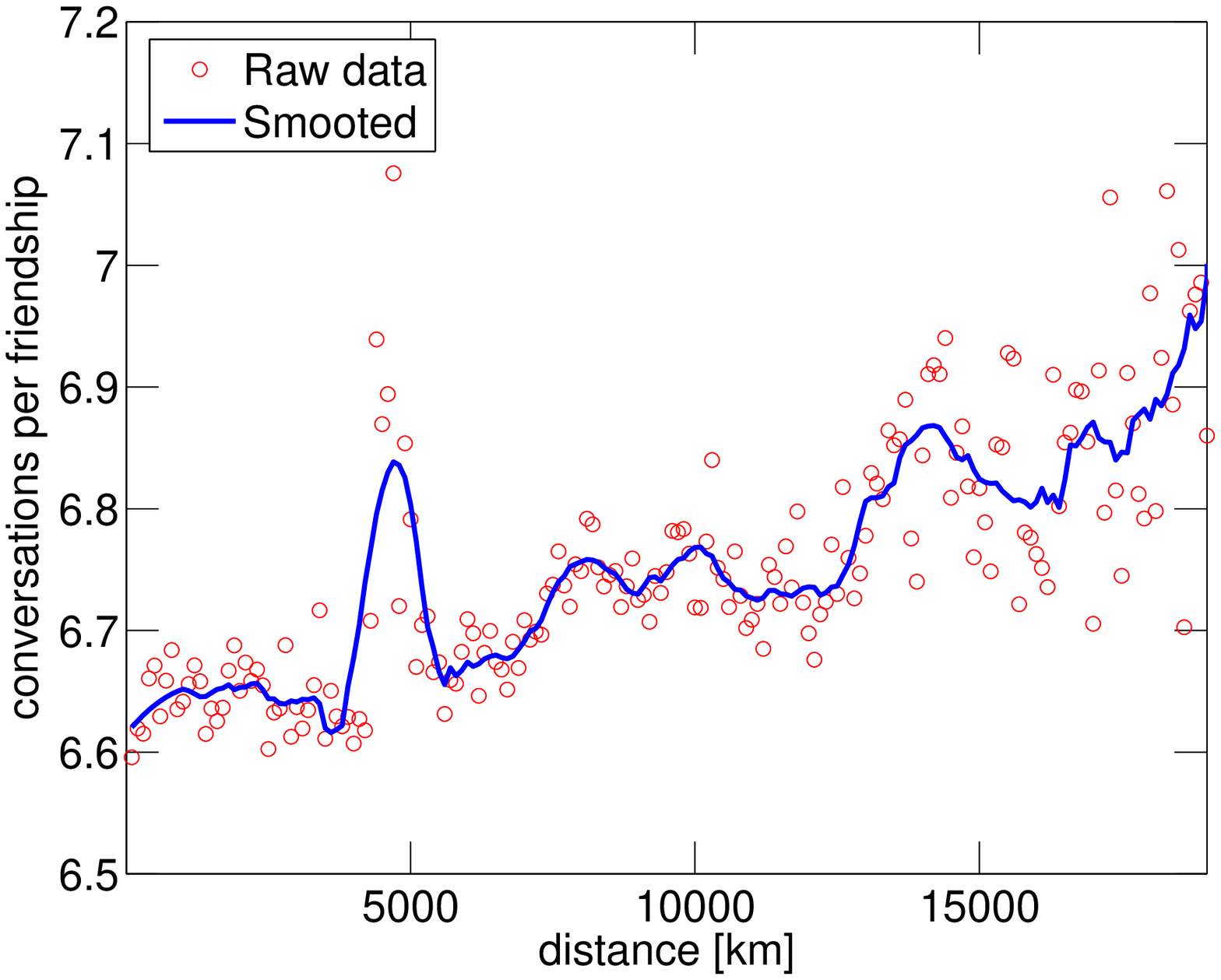} &
  \includegraphics[width=0.45\textwidth]{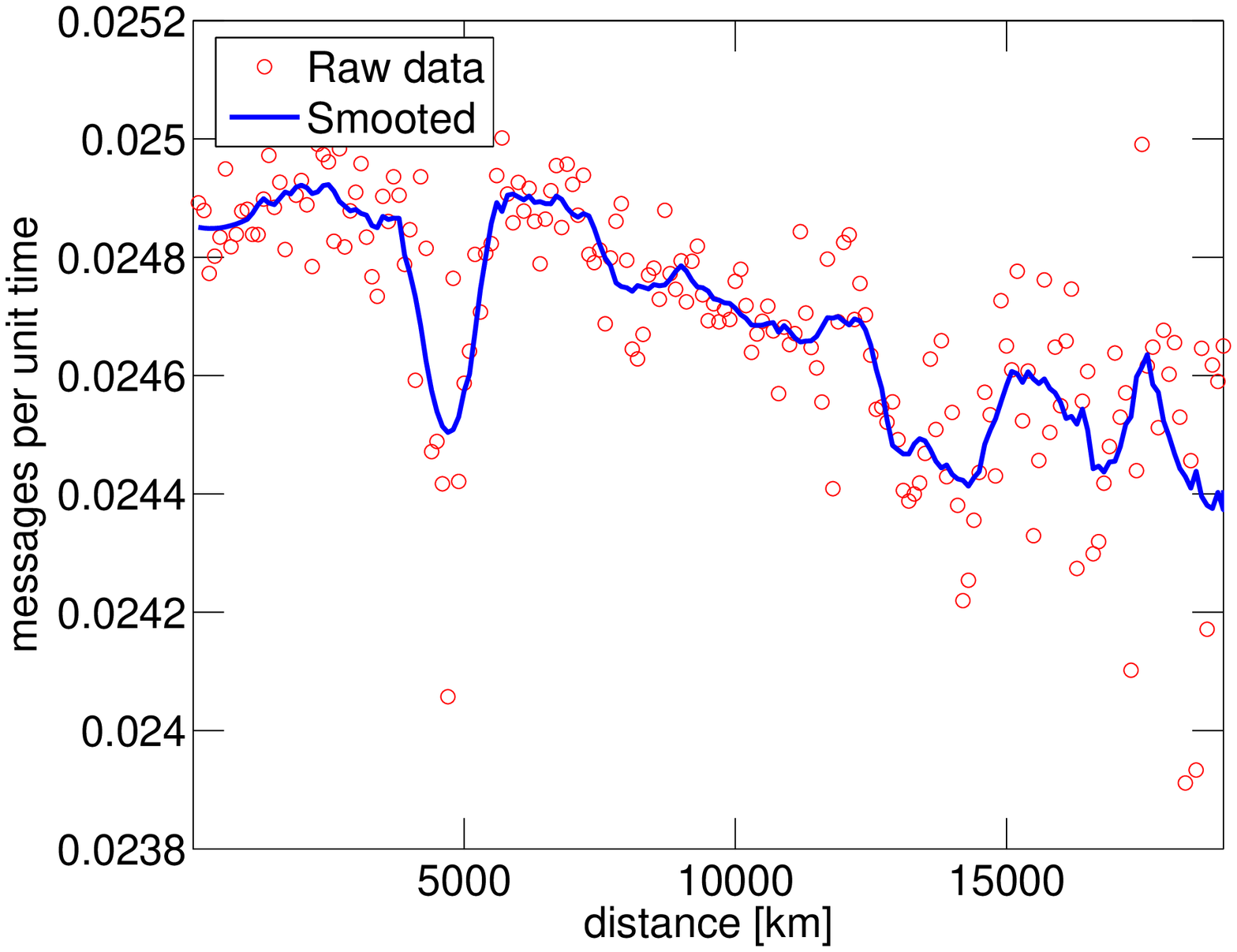} \\
  (e) Conversations per link & (f) Messages per unit time \\
  \end{tabular}
  \end{center}
  \caption{Communication with the distance. (a) Number of links (pairs
  of people that communicate) with the distance. (b) Number of
  conversations between people at particular distance. (c) Average
  conversation duration. (d) Number of exchanged messages per
  conversation.
  (e) Number of conversations per link (per pair of communicating users).
  (f) Number of exchanged messages per unit time.}
  \label{fig:commDist}
\end{figure}

Figure~\ref{fig:commDist} plots the relation between communication and
distance. Figure~\ref{fig:commDist}(a) shows the distribution of the number
of conversations between conversants at distance $l$. We found that the
number of conversations decreases with distance. However, we observe a peak
at a distance of approximately 500 kilometers. The other peaks and drops may
reveal geographical features. For example, a significant drop in
communication at distance of 5,000 km (3,500 miles) may reflect the width of
the Atlantic ocean or the distance between the east and west coasts of the
United States. The number of links rapidly decreases with distance. This
finding suggests that users may use Messenger mainly for communications with
others within a local context and environment. We found that the number of
exchanged messages and conversation lengths do not increase with distance
(see plots (b)--(d) and (f) of Figure~\ref{fig:commDist}). Conversation
duration decreases with the distance, while the number of exchanged messages
remains constant before decreasing slowly. Figure~\ref{fig:commDist}(f) shows
the communications per link versus the distance among participants. The plot
shows that longer links, {\em i.e.}, connections between people who are
farther apart, are more frequently used than shorter links. We interpret this
finding to mean that people who are farther apart use Messenger more
frequently to communicate.

In summary, we observe that the total number of links and associated
conversations decreases with increasing distance among participants. The same
is true for the duration of conversations, the number of exchanged messages
per conversation, and the number of exchanged messages per unit time.
However, the number of times a link is used tends to increase with the
distance among users. This suggests that people who are farther apart tend to
converse with IM more frequently, which perhaps takes the place of more
expensive long-distance voice telephony; voice might be used more frequently
in lieu of IM for less expensive local communications.

\section{Homophily of communication}
\label{sec:homophily}

We performed several experiments to measure the level at which people tend to
communicate with similar people. First, we consider all 1.3 billion pairs of
people who exchanged at least one message in June 2006, and calculate the
similarity of various user demographic attributes. We contrast this with the
similarity of pairs of users selected via uniform random sampling across 180
million users. We consider two measures of similarity: the correlation
coefficient and the probability that users have the same attribute value,
{\em e.g.}, that users come from the same countries.

\begin{table}
  \begin{center}
  \begin{tabular}{l|c|c||c|c}
    & \multicolumn{2}{c||}{Correlation} & \multicolumn{2}{c}{Probability} \\
    Attribute & Rnd & Comm & Rnd & Comm\\ \hline
    Age & -0.0001 & 0.297 & 0.030 & 0.162 \\
    Gender & 0.0001 & -0.032 & 0.434 & 0.426 \\
    ZIP & -0.0003 & 0.557 & 0.001 & 0.23 \\
    County & 0.0005 & 0.704 & 0.046 & 0.734 \\
    Language & -0.0001 & 0.694 & 0.030 & 0.798 \\
  \end{tabular}
  \end{center}
  \caption{Correlation coefficients
  and probability of users sharing an attribute for random pairs of people versus for pairs of people who communicate.}
  \label{tab:homoCorr}
  \label{tab:homoProb}
\end{table}

Table~\ref{tab:homoCorr} compares correlation coefficients of various user
attributes when pairs of users are chosen uniformly at random with
coefficients for pairs of users who communicate. We can see that attributes
are not correlated for random pairs of people, but that they are highly
correlated for users who communicate. As we noted earlier, gender and
communication are slightly negatively correlated; people tend to communicate
more with people of the opposite gender.

Another method for identifying association is to measure the probability that
a pair of users will show an exact match in values of an attribute, {\em
i.e.}, identifying whether two users come from the same country, speak the
same language, etc. Table~\ref{tab:homoProb} shows the results for the
probability of users sharing the same attribute value. We make similar
observations as before. People who communicate are more likely to share
common characteristics, including age, location, language, and they are less
likely to be of the same gender. We note that the most common attribute of
people who communicate is language. On the flip side, the amount of
communication tends to decrease with increasing user dissimilarity. This
relationship is highlighted in Figure~\ref{fig:commDist}, which shows how
communication among pairs of people decreases with distance.

\begin{figure}
  \begin{center}
  \begin{tabular}{cc}
  \includegraphics[width=0.45\textwidth]{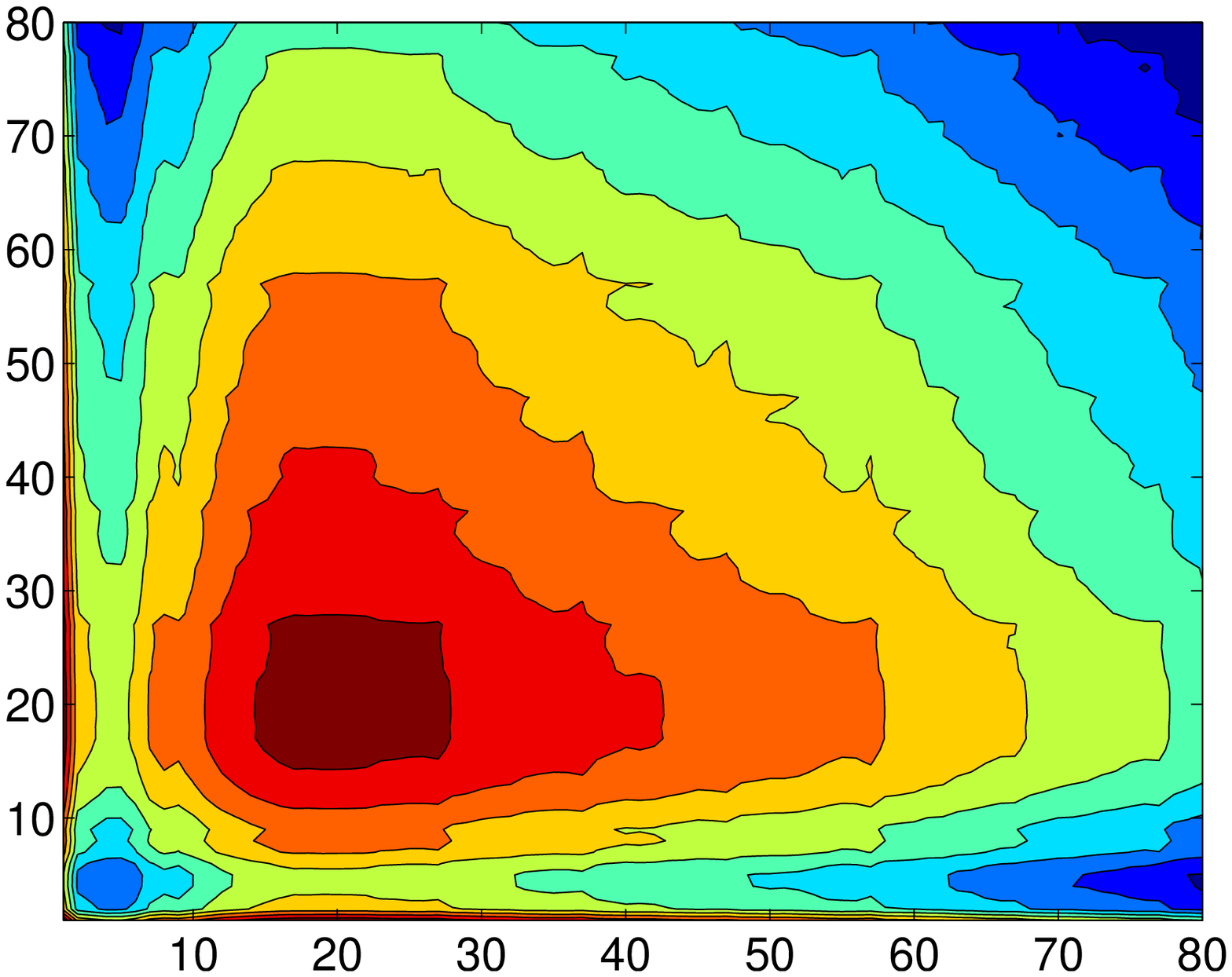} &
  \includegraphics[width=0.45\textwidth]{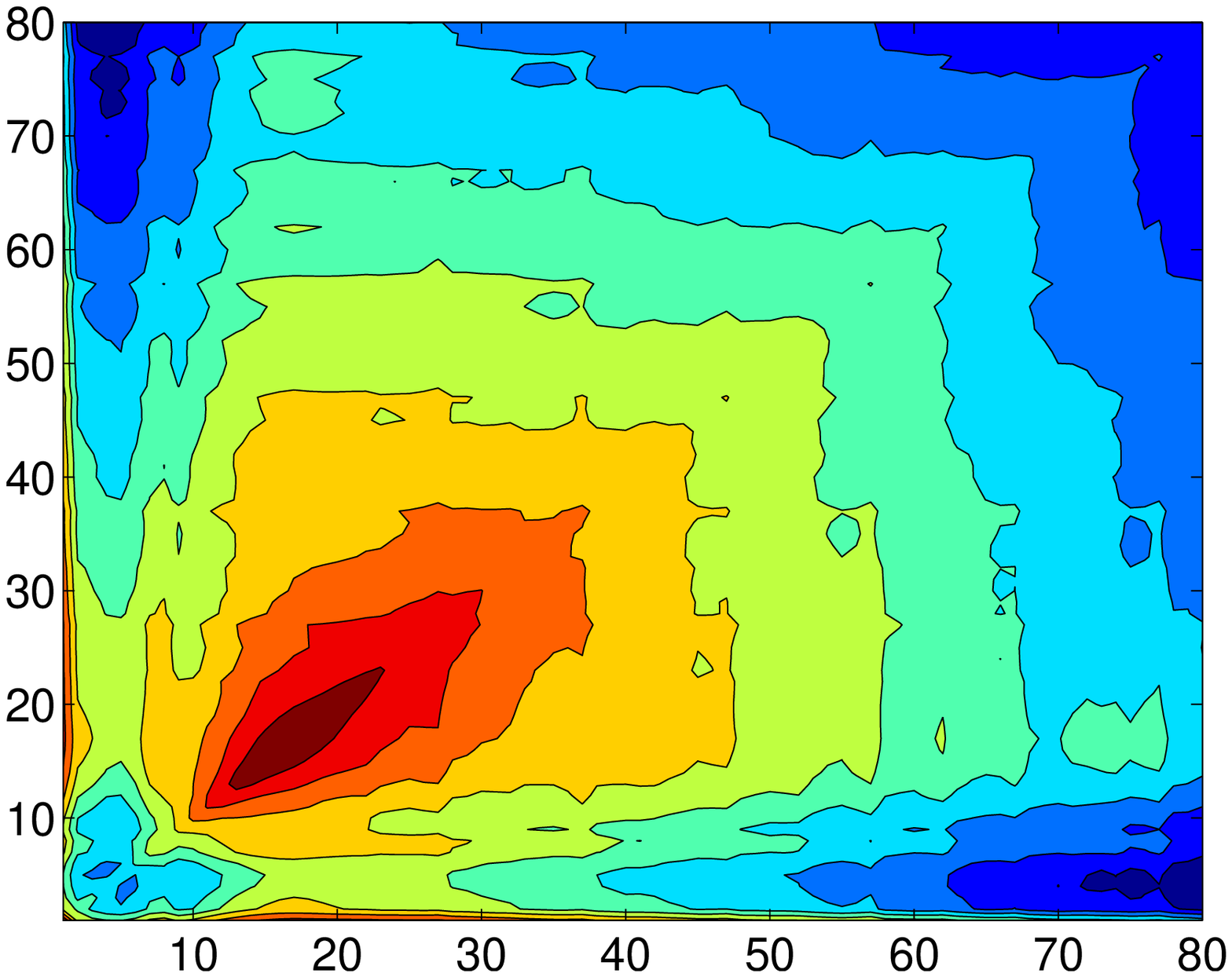} \\
  (a) Random & (b) Communicate \\
  \end{tabular}
  \end{center}
 \caption{Numbers of pairs of people of different ages.
 (a) Randomly selected pairs of people; (b) people who
 communicate. Correlation between age and communication is captured by the diagonal trend.}
  \label{fig:homoAgeComm}
\end{figure}

Figure~\ref{fig:homoAgeComm} further illustrates the results displayed in
Table~\ref{tab:homoCorr}, where we randomly sample pairs of users from the
Messenger user base, and then plot the distribution over reported ages. As
most of the population comes from the age group 10--30, the distribution of
random pairs of people reaches the mode at those ages but there is no
correlation. Figure~\ref{fig:homoAgeComm}(b) shows the distribution of ages
over the pairs of people who communicate. Note the correlation, as
represented by the diagonal trend on the plot, where people tend to
communicate more with others of a similar age.

\begin{figure}
\begin{center}
  \begin{tabular}{cc}
  \includegraphics[width=0.45\textwidth]{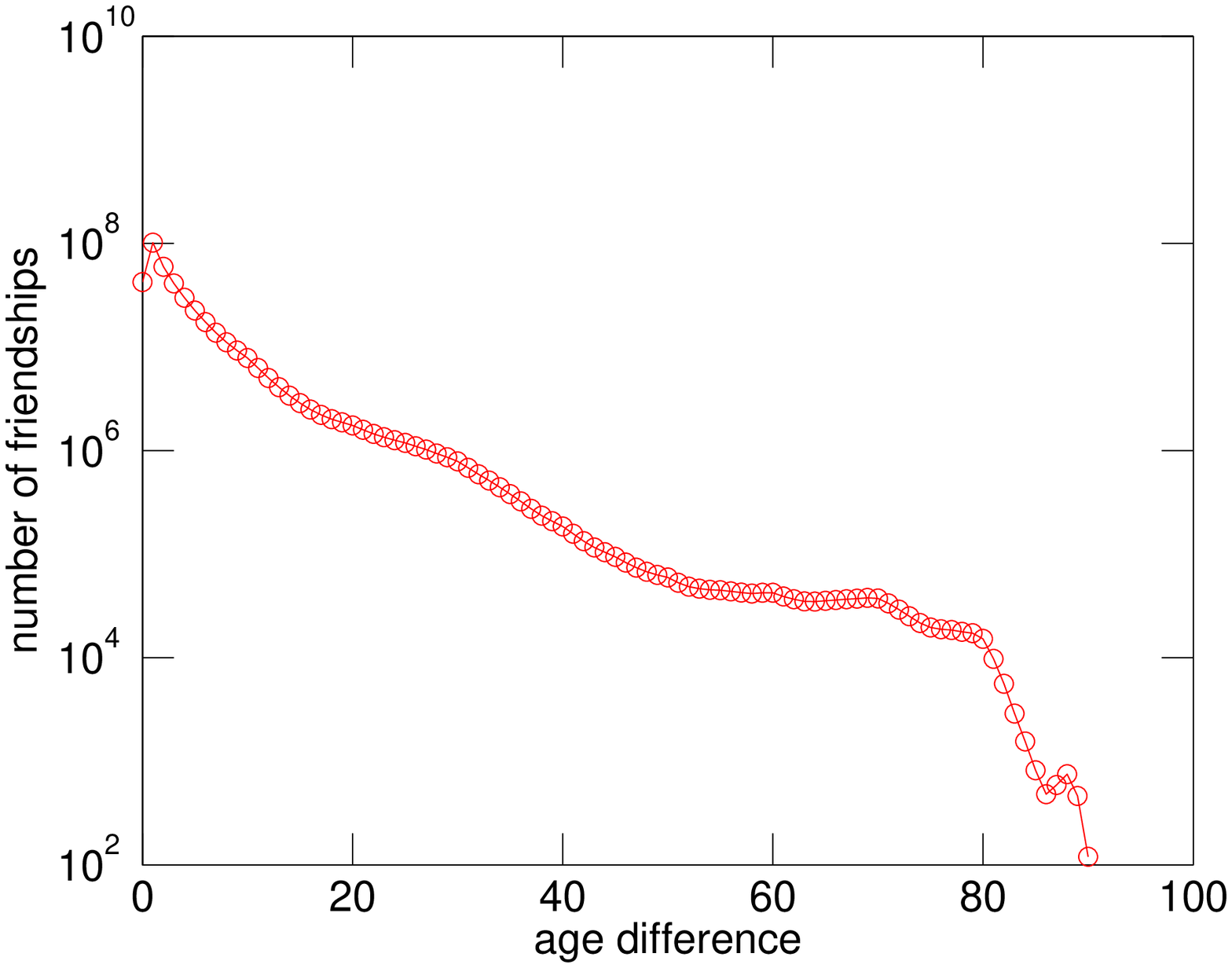} &
  \includegraphics[width=0.45\textwidth]{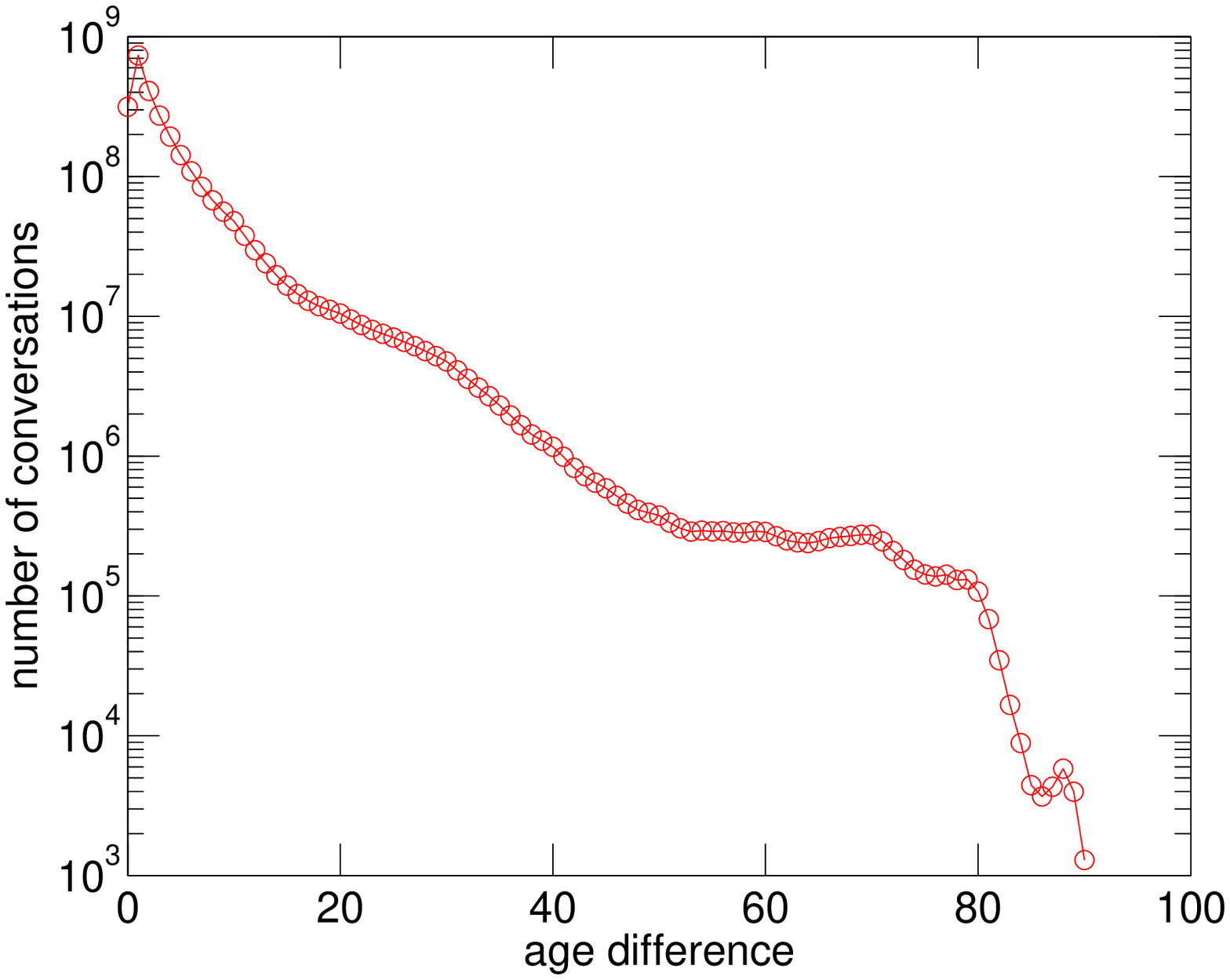} \\
  (a) Number of links & (b) Number of conversations \\
  \includegraphics[width=0.45\textwidth]{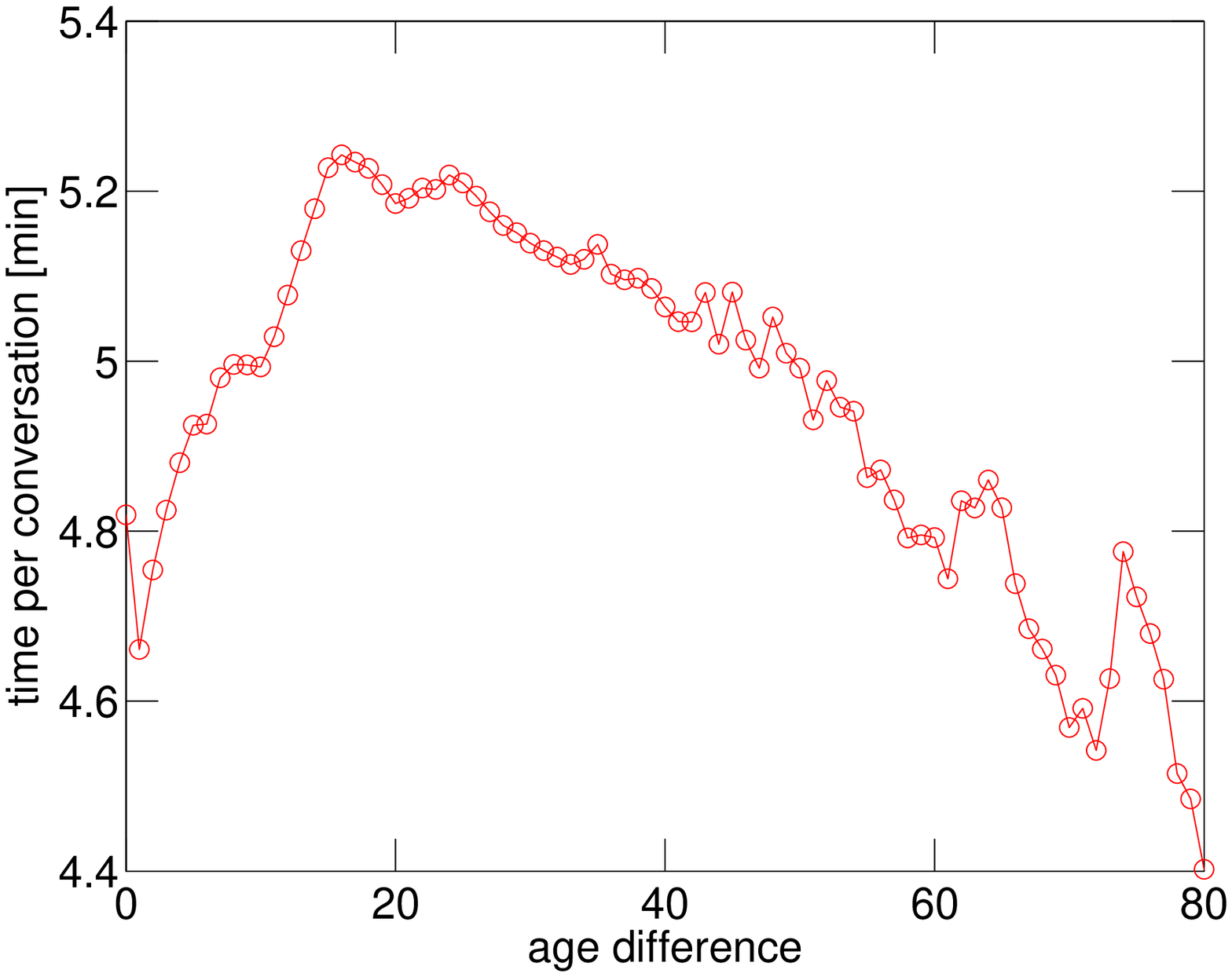} &
  \includegraphics[width=0.45\textwidth]{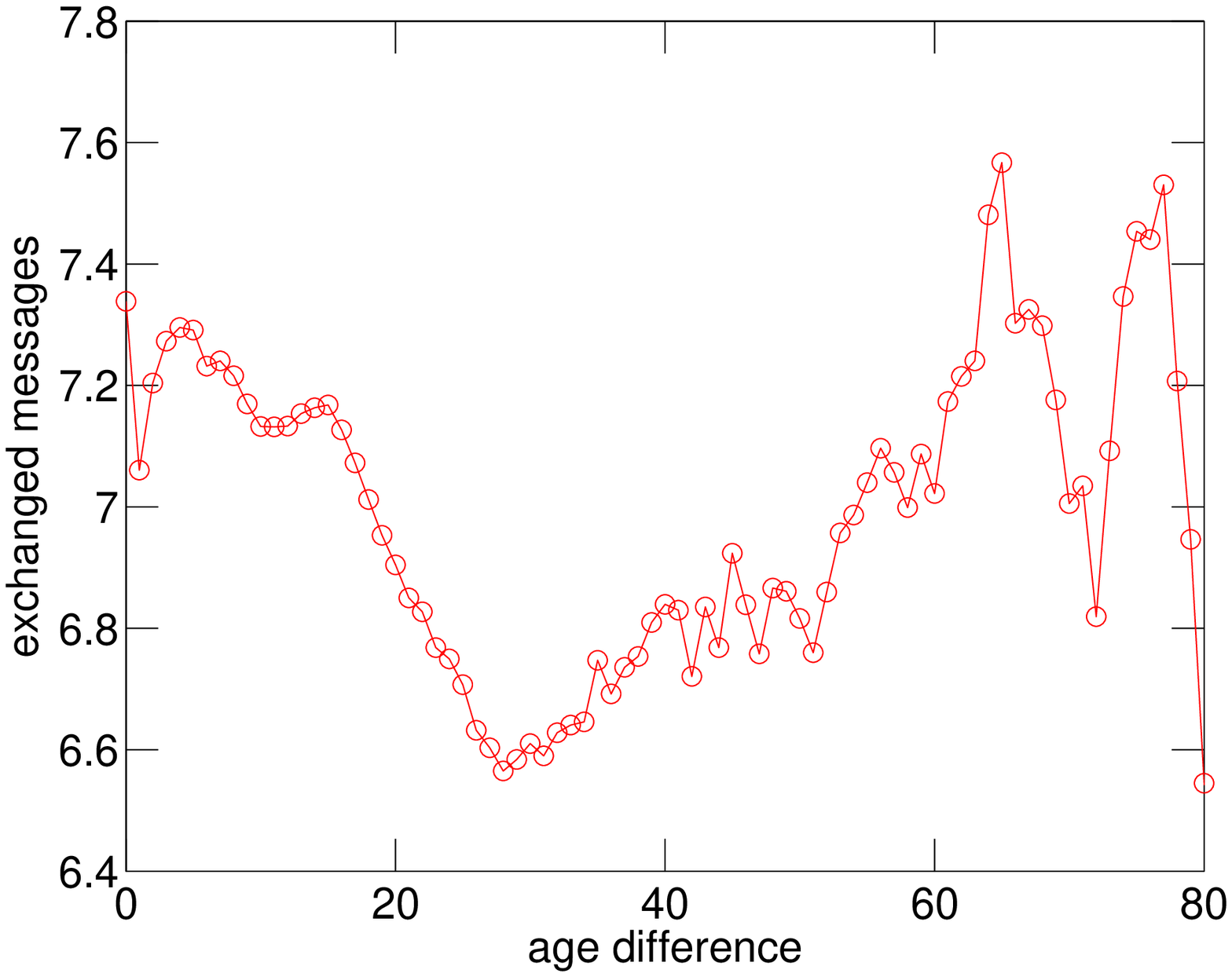} \\
  (c) Conversation duration & (d) Exchanged messages \\
  \includegraphics[width=0.45\textwidth]{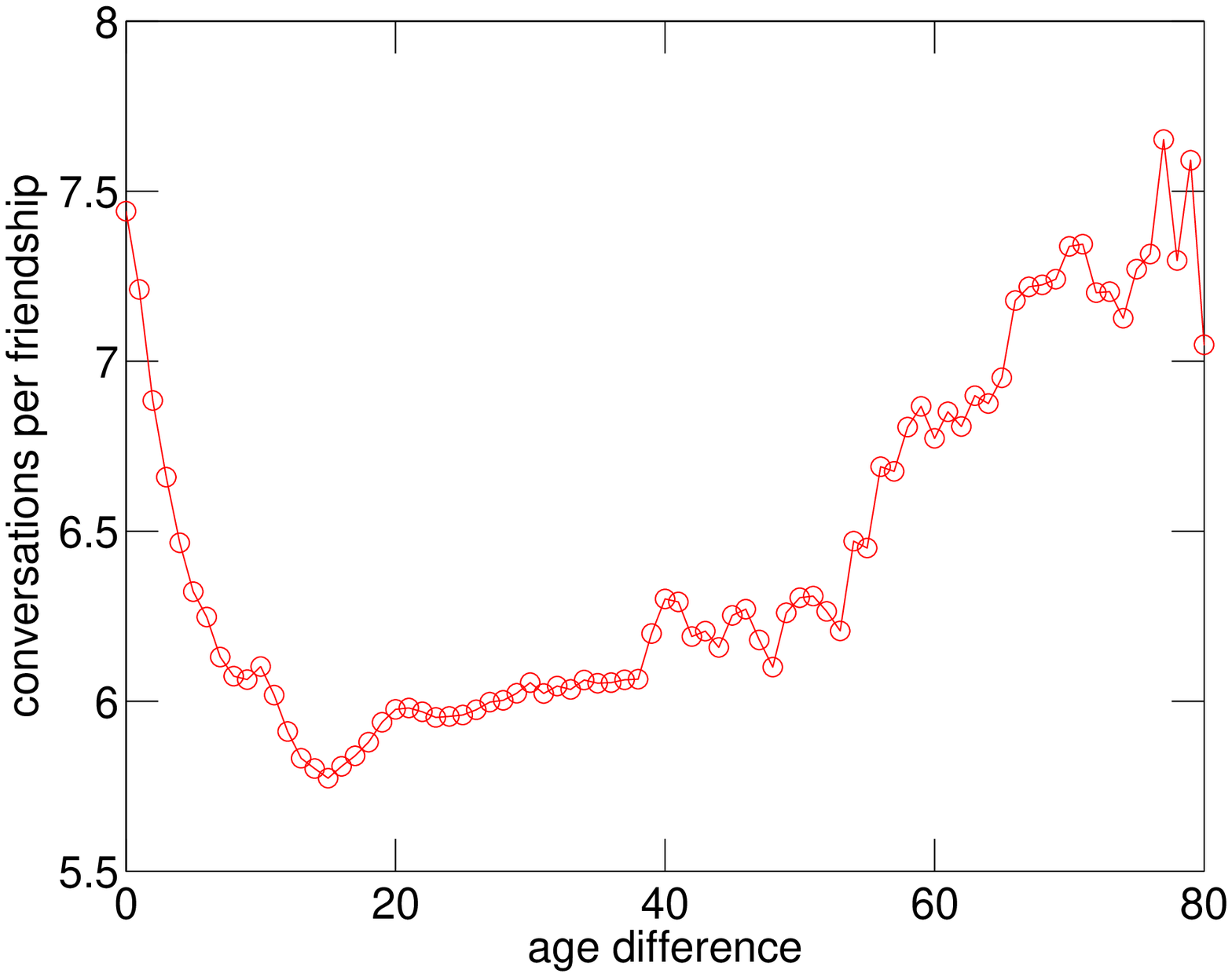} &
  \includegraphics[width=0.45\textwidth]{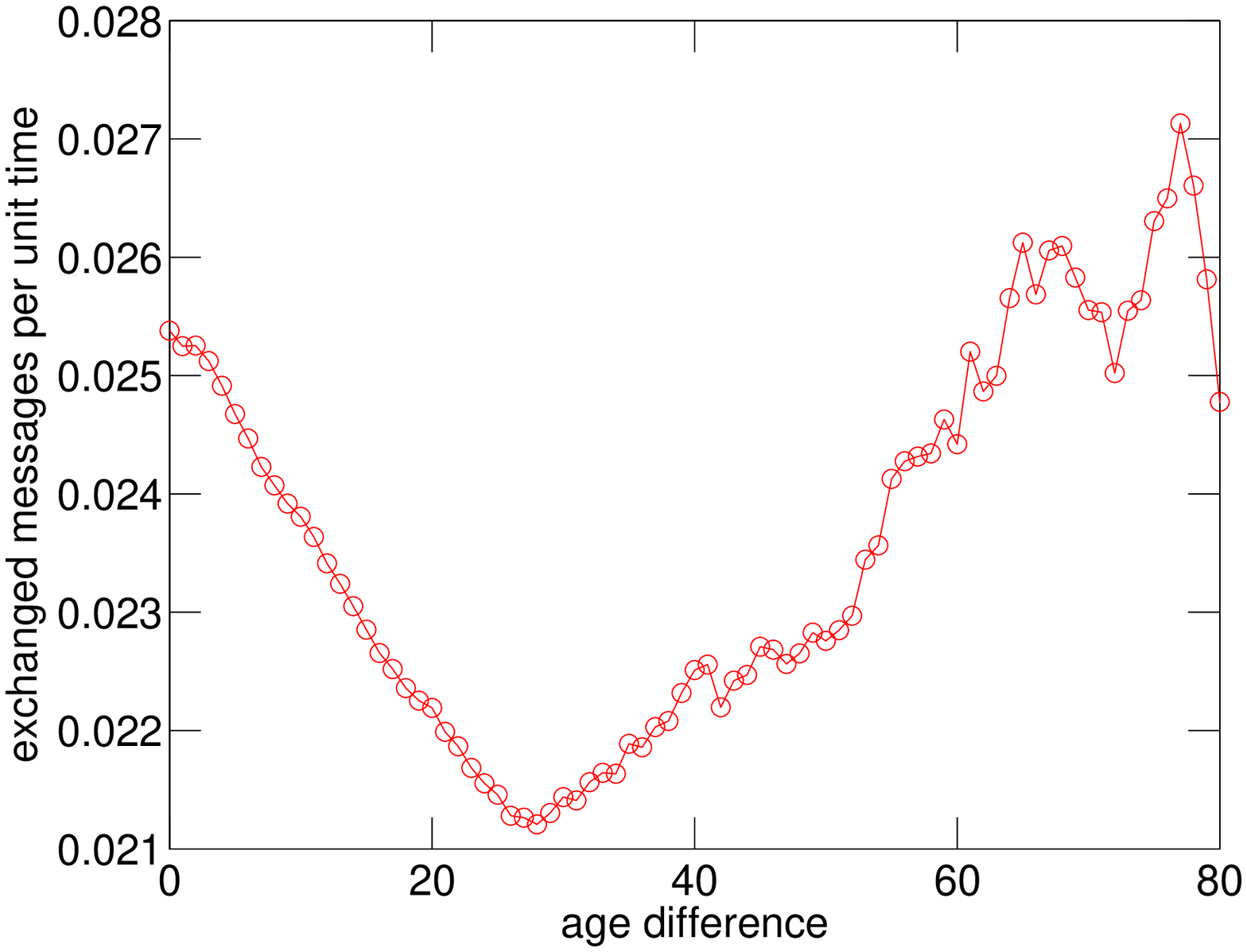} \\
  (e) Conversations per link & (f) Messages per unit time \\
  \end{tabular}
  \caption{Communication characteristics with age difference between the
  users. (a) Number of links (pairs communicating) with the age
  difference. (b) Number of conversations. (c) Average conversation
  duration with the age difference. (d) Average number of exchanged
  messages per conversation as a function of the age difference between
  the users. (e) Number of conversations per link in the observation
  period with the age difference. (f) Number of exchanged messages per
  unit time as a function of age difference between the users.}
  \label{fig:commAgeDiff}
\end{center}
\end{figure}

Next, we further explore communication patterns by the differences in the
reported ages among users. Figure~\ref{fig:commAgeDiff}(a) plots the number
links in the communication network vs. the age difference of the
communicating pair of users. Similarly, Figure~\ref{fig:commAgeDiff}(b) plots
on a log-linear scale the number of conversations in the social network with
participants of varying age differences. Again we see that links and
conversations are strongly correlated with the age differences among
participants. Figure~\ref{fig:commAgeDiff}(c) shows the average conversation
duration with the age difference among the users. Interestingly, the mean
conversation duration peaks at an age difference of 20 years between
participants. We speculate that the peak may correspond roughly to the gap
between generations.

The plots reveal that there is strong homophily in the communication network
for age; people tend to communicate more with people of similar reported age.
This is especially salient for the number of buddies and conversations among
people of the same ages. We also observe that the links between people of
similar attributes are used more often, to interact with shorter and more
intense (more exchanged messages) communications. The intensity of
communication decays linearly with the difference in age. In contrast to
findings of previous studies, we observe that the number of cross-gender
communication links follows a random chance. However, cross-gender
communication takes longer and is faster paced as it seems that people tend
to pay more attention when communicating with the opposite sex.

Recently, using the data we generated, Singla and Richardson further
investigated the homophily within the Messenger network and found that people
who communicate are also more likely to search the web for content on similar
topics~\cite{singla08correlation}.

\section{The communication network}
\label{sec:network}

So far we have examined communication patterns based on pairwise
communications. We now create a more general communication network from the
data. Using this network, we can examine the typical {\em social distance}
between people, {\em i.e.}, the number of links that separate a random pair
of people. This analysis seeks to understand how many people can be reached
within certain numbers of hops among people who communicate. Also, we test
the transitivity of the network, {\em i.e.}, the degree at which pairs with a
common friend tend to be connected.

We constructed a graph from the set of all two-user conversations, where each
node corresponds to a person and there is an undirected edge between a pair
of nodes if the users were engaged in an active conversation during the
observation period (users exchanged at least 1 message). The resulting
network contains 179,792,538 nodes, and 1,342,246,427 edges. Note that this
is not simply a {\em buddy network}; we only connect people who are buddies
{\em and} have communicated during the observation period.

\begin{figure}
\begin{center}
    \begin{tabular}{cc}
    \includegraphics[width=0.45\textwidth]{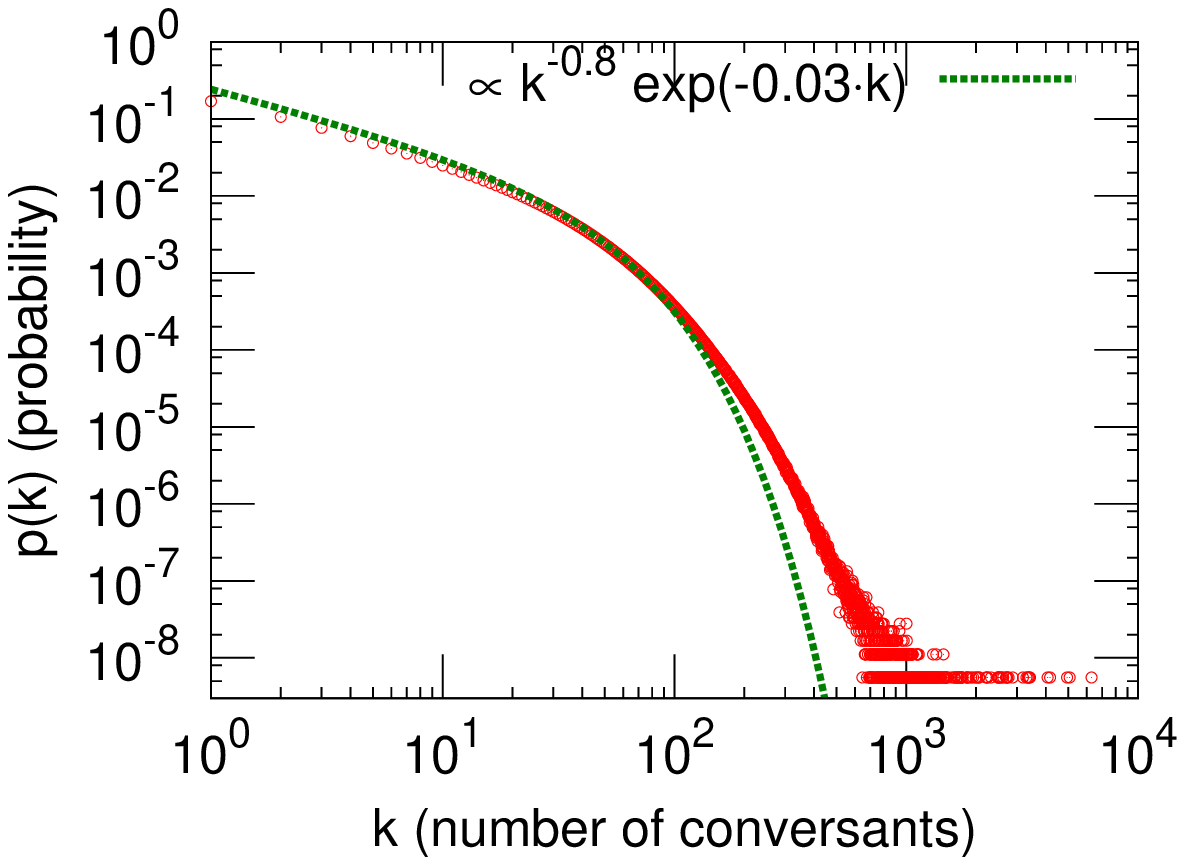} &
    \includegraphics[width=0.45\textwidth]{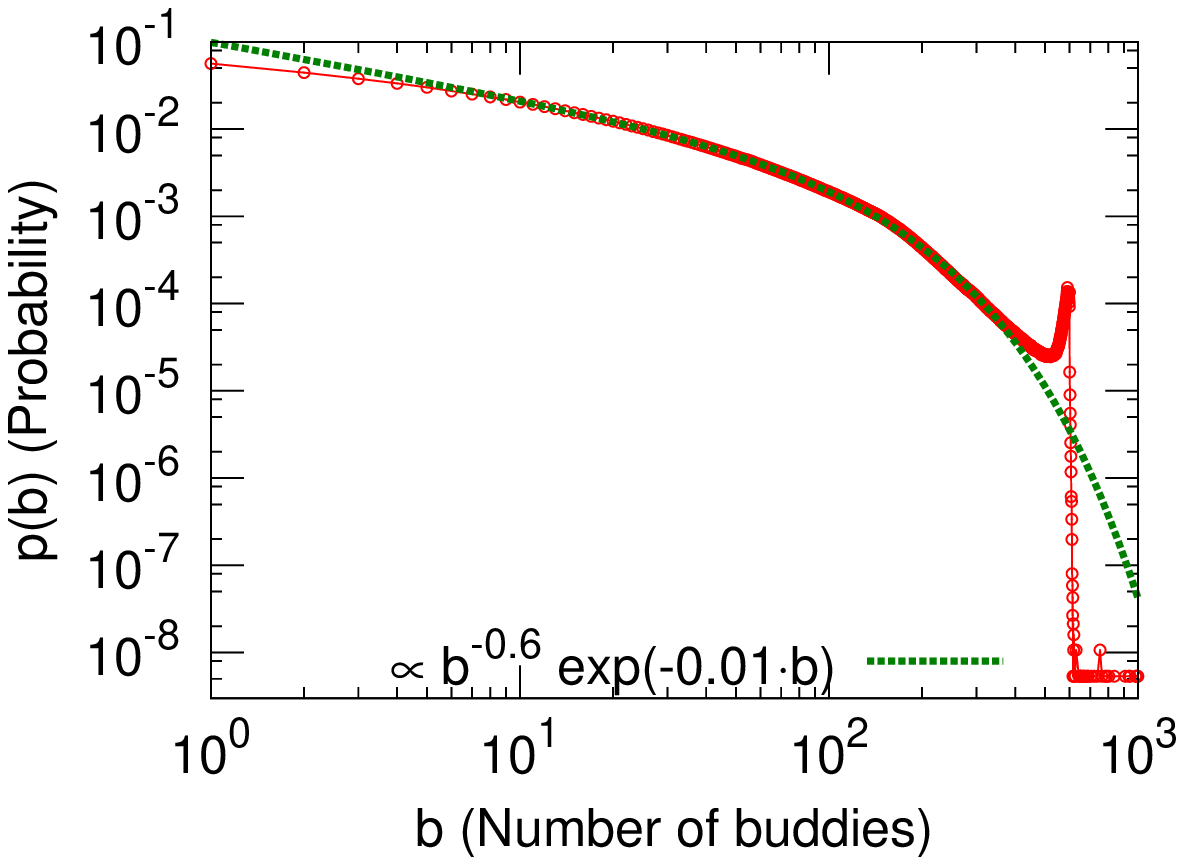} \\
    (a) Communication & (b) Buddies \\
    \end{tabular}
  \caption{(a) Degree distribution of communication network (number of people with whom a person communicates).
    (b) Degree distribution of the buddy network (length of the contact list).}
  \label{fig:commNet-Deg}
  \label{fig:buddiesA}
\end{center}
\end{figure}

\begin{figure}
\begin{center}
    \begin{tabular}{cc}
    \includegraphics[width=0.45\textwidth]{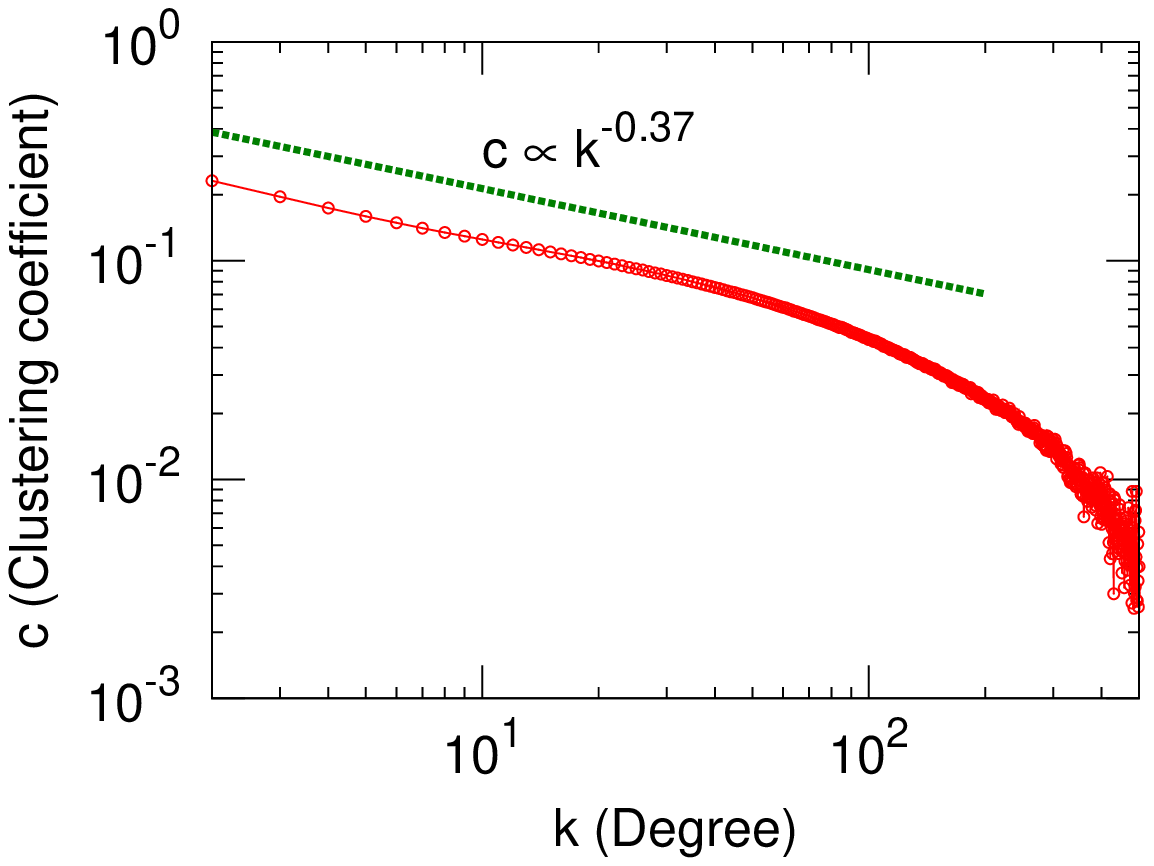} &
    \includegraphics[width=0.45\textwidth]{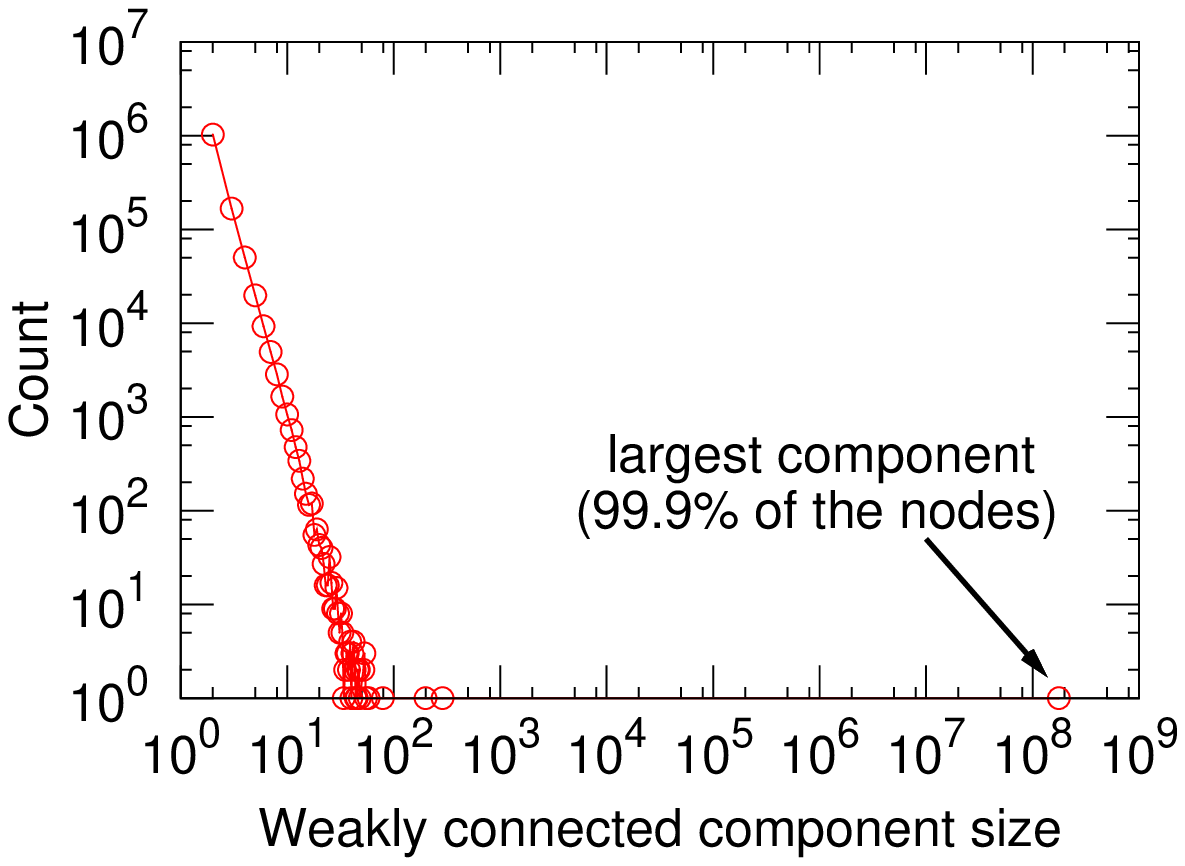} \\
    (a) Clustering & (b) Components \\
    \end{tabular}
  \caption{(a) Clustering coefficient; (b) distribution of connected
  components. 99.9\% of the nodes belong to the largest connected component. }
  \label{fig:commNet-Wcc}
  \label{fig:commNet-CCf}
\end{center}
\end{figure}

Figures~\ref{fig:commNet-Deg}--\ref{fig:commNet-Wcc} show the structural
properties of the communication network. The network degree distribution
shown in Figure~\ref{fig:commNet-Deg}(a) is heavy tailed but does not follow
a power-law distribution. Using maximum likelihood estimation, we fit a
power-law with exponential cutoff $p(k) \propto k^{-a} e^{-b k}$ with fitted
parameter values $a=0.8$ and $b=0.03$. We found a strong cutoff parameter and
low power-law exponent, suggesting a distribution with high variance.

Figure~\ref{fig:buddiesA}(b) displays the degree distribution of a buddy
graph.  We did not have access to the full buddy network; we only had access
to data on the length of the user contact list which allowed us to create the
plot. We found a total of 9.1 billion buddy edges in the graph with 49
buddies per user. We fit the data with a power-law distribution with
exponential cutoff and identified parameters of $a=0.6$ and $b=0.01$. The
power-law exponent now is even smaller. This model described the data well.
We note a spike at 600 which is the limit on the maximal number of buddies
imposed by the Messenger software client. The maximal number of buddies was
increased to 300 from 150 in March 2005, and was later raised to 600. With
the data from June 2006, we see only the peak at 600, and could not identify
bumps at the earlier constraints.

Social networks have been found to be highly transitive, {\em i.e.}, people
with common friends tend to be friends themselves. The clustering
coefficient~\cite{watts98smallworld} has been used as a measure of
transitivity in the network. The measure is defined as the fraction of
triangles around a node of degree $k$~\cite{watts98smallworld}.
Figure~\ref{fig:commNet-CCf}(a) displays the clustering coefficient versus
the degree of a node for Messenger. Previous results on measuring the web
graph as well as theoretical analyses show that the clustering coefficient
decays as $k^{-1}$ (exponent $-1$) with node degree
$k$~\cite{ravasz03hierar}. For the Messenger network, the clustering
coefficient decays very slowly with exponent $-0.37$ with the degree of a
node and the average clustering coefficient is 0.137. This result suggests
that clustering in the Messenger network is much higher than expected---that
people with common friends also tend to be connected.
Figure~\ref{fig:commNet-Wcc}(b) displays the distribution of the connected
components in the network. The giant component contains 99.9\% of the nodes
in the network against a background of small components, and the distribution
follows a power law.

\begin{figure}[t]
\begin{center}
    \begin{tabular}{cc}
    \includegraphics[width=0.45\textwidth]{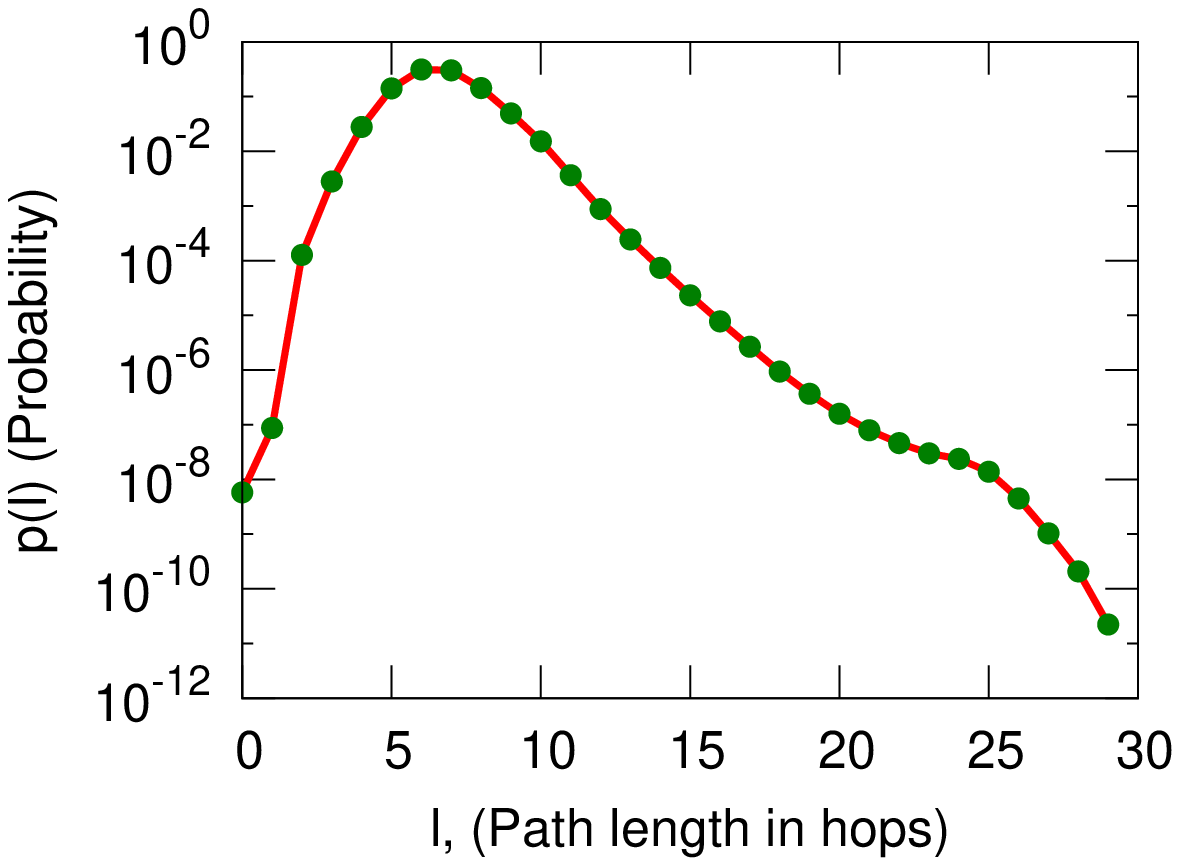} &
    \includegraphics[width=0.45\textwidth]{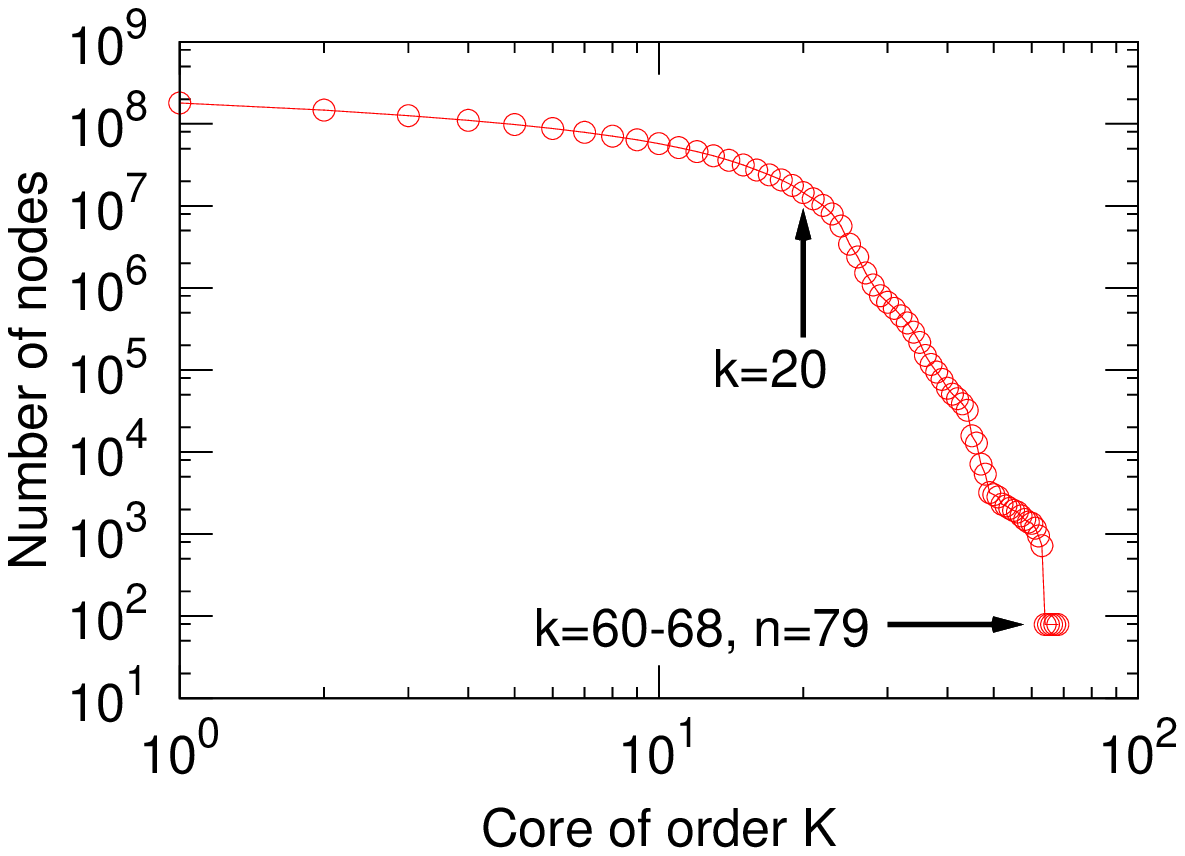} \\
    (a) Diameter & (b) $k$-cores \\
    \end{tabular}
  \caption{
  (a) Distribution over the shortest path lengths. Average
  shortest path has length 6.6, the distribution reaches the
  mode at 6 hops, and the 90\% effective diameter is 7.8;
  (b) distribution of sizes of cores of order $k$.}
  \label{fig:commNet-Diam}
  \label{fig:commNet-cores}
\end{center}
\end{figure}

\subsection{How small is the small world?}

Messenger data gives us a unique opportunity to study distances in the social
network. To our knowledge, this is the first time a planetary-scale social
network has been available to validate the well-known ``6 degrees of
separation'' finding by Travers and Milgram~\cite{milgram69smallworld}.  The
earlier work employed a sample of 64 people and found that the average number
of hops for a letter to travel from Nebraska to Boston was 6.2 (mode 5,
median 5), which is popularly known as the ``6 degrees of separation'' among
people. We used a population sample that is more than two million times
larger than the group studied earlier and confirmed the classic finding.

Figure~\ref{fig:commNet-Diam}(a) displays the distribution over the shortest
path lengths. To approximate the distribution of the distances, we randomly
sampled 1000 nodes and calculated for each node the shortest paths to all
other nodes. We found that the distribution of path lengths reaches the mode
at 6 hops and has a median at 7. The average path length is 6.6. This result
means that a random pair of nodes in the Messenger network is 6.6 hops apart
on the average, which is half a link longer than the length measured by
Travers and Milgram. The 90th percentile (effective
diameter~\cite{tauro01topology}) of the distribution is 7.8. 48\% of nodes
can be reached within 6 hops and 78\% within 7 hops. So, we might say that,
via the lens provided on the world by Messenger, we find that there are about
``7 degrees of separation'' among people. We note that long paths, {\em
i.e.}, nodes that are far apart, exist in the network; we found paths up to a
length of 29.

\subsection{Network cores}

\begin{figure}
\begin{center}
    \includegraphics[width=0.5\textwidth]{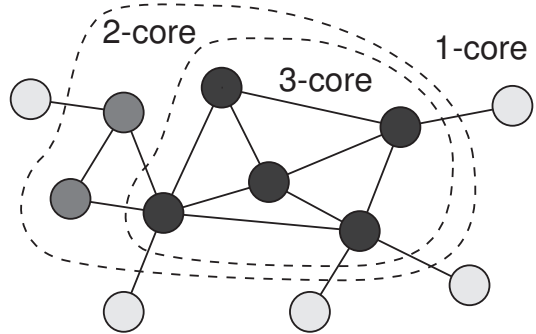}
  \caption{$k$-core decomposition of a small graph. Nodes contained in
  each closed line belong to a given $k$-core. Inside each $k$-core all
  nodes have degree larger than $k$ (after removing all nodes with
  degree less than $k$.}
  \label{fig:k-core}
\end{center}
\end{figure}

We further study connectivity of the communication network by examining the
$k$-cores~\cite{batagelj05cores} of the graph. The concept of $k$-core is a
generalization of the giant connected component. The $k$-core of a network is
a set of vertices $K$, where each vertex in $K$ has at least $k$ edges to
other vertices in $K$ (see Figure~\ref{fig:k-core}). The distribution of
$k$-core sizes gives us an idea of how quickly the network shrinks as we move
towards the core.

The $k$-core of a graph can be obtained by deleting from the network all
vertices of degree less than $k$. This process will decrease degrees of some
non-deleted vertices, so more vertices will have degree less than $k$. We
keep pruning vertices until all remaining vertices have degree of at least
$k$. We call the remaining vertices a $k$-core.

Figure~\ref{fig:commNet-cores} plots the number of nodes in a core of order
$k$. We note that the core sizes are remarkably stable up to a value of
$k\approx 20$; the number of nodes in the core drops for only an order of
magnitude. After $k>20$, the core size rapidly drops. The central part of the
communication network is composed of 79 nodes, where each of them has more
than 68 edges inside the set. The structure of the Messenger communication
network is quite different from the Internet graph; it has been
observed~\cite{alvarez05cores} that the size of a $k$-core of the Internet
decays as a power-law with $k$. Here we see that the core sizes remains very
stable up to a degree $\approx 20$, and only then start to rapidly degrease.
This means that the nodes with degrees of less than 20 are on the fringe of
the network, and that the core starts to rapidly decrease as nodes of degree
20 or more are deleted.

\subsection{Strength of the ties}

It has been observed by Albert et al.~\cite{albert00error} that many
real-world networks are robust to node-level changes or {\em attacks}.
Researchers have showed that networks like the World Wide Web, Internet, and
several social networks display a high degree of robustness to random node
removals, {\em i.e.}, one has to remove many nodes chosen uniformly at random
to make the network disconnected. On the contrary, targeted attacks are very
effective. Removing a few high degree nodes can have a dramatic influence on
the connectivity of a network.

Let us now study how the Messenger communication network is decomposed when
``strong,'' {\em i.e.}, heavily used, edges are removed from the network. We
consider several different definitions of ``heavily used,'' and measure the
types of edges that are most important for network connectivity. We note that
a similar experiment was performed by Shi et al~\cite{shi06strongties} in the
context of a small IM buddy network. The authors of the prior study took the
number of common friends at the ends of an edge as a measure of the link
strength. As the number of edges here is too large (1.3 billion) to remove
edges one by one, we employed the following procedure: We order the nodes by
decreasing value per a measure of the {\em intensity of engagement} of users;
we then delete nodes associated with users in order of decreasing measure and
we observe the evolution of the properties of the communication network as
nodes are deleted.

We consider the following different measures of engagement:
\begin{itemize}
	\item Average sent: The average number of sent messages per user's
conversation
  \item Average time: The average duration of user's conversations
	\item Links: The number of links of a user (node degree), {\em i.e.},
number of different people he or she exchanged messages with
	\item Conversations: The total number of conversations of a user in the
observation period
	\item Sent messages: The total number of sent messages by a user in the
observation period
  \item Sent per unit time: The number of sent messages per unit time of
      a conversation
	\item Total time: The total conversation time of a user in the
observation period
\end{itemize}

At each step of the experiment, we remove 10 million nodes in order of the
specific measure of engagement being studied. We then determine the relative
size of the largest connected component, {\em i.e.}, given the network at
particular step, we find the fraction of the nodes belonging to the largest
connected component of the network.

\begin{figure}
\begin{center}
    \includegraphics[width=0.9\textwidth]{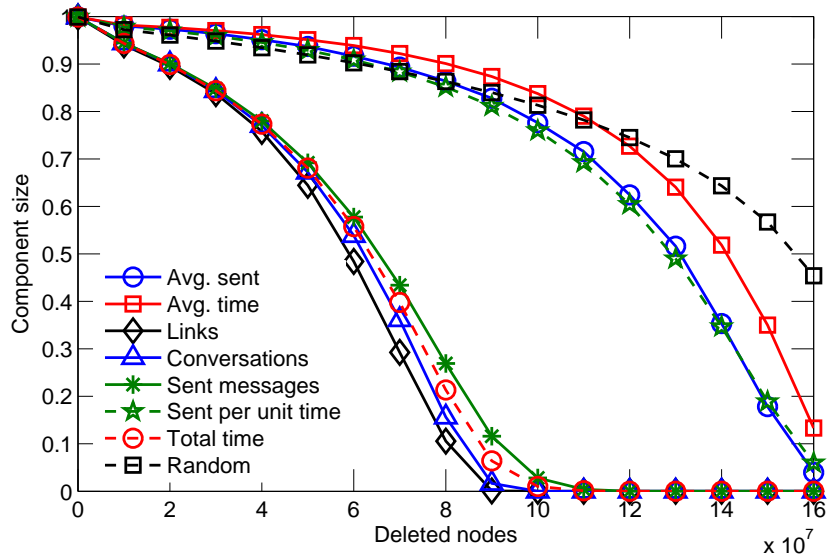}
  \caption{Relative size of the largest
  connected component as a function of number of nodes removed.}
  \label{fig:commNet-delNodesWcc}
\end{center}
\end{figure}

Figure~\ref{fig:commNet-delNodesWcc} plots the evolution of the fraction of
nodes in the largest connected component with the number of deleted nodes. We
plot a separate curve for each of the seven different measures of engagement.
For comparison, we also consider the random deletion of the nodes.

The decomposition procedure highlighted two types of dynamics of network
change with node removal. The size of the largest component decreases rapidly
when we use as measures of engagement the number of links, number of
conversations, total conversation time, or number of sent messages. In
contrast, the size of the largest component decreases very slowly when we use
as a measure of engagement the average time per conversation, average number
of sent messages, or number of sent messages per unit time. We were not
surprised to find that the size of the largest component size decreases most
rapidly when nodes are deleted in order of the decreasing number of links
that they have, {\em i.e.}, the number of people with whom a user at a node
communicates. Random ordering of the nodes shrinks the component at the
slowest rate. After removing 160 million out of 180 million nodes with the
random policy, the largest component still contains about half of the nodes.
Surprisingly, when deleting up to 100 million nodes, the average time per
conversation measure shrinks the component even more slowly than the random
deletion policy.

\begin{figure}
\begin{center}
    \includegraphics[width=0.9\textwidth]{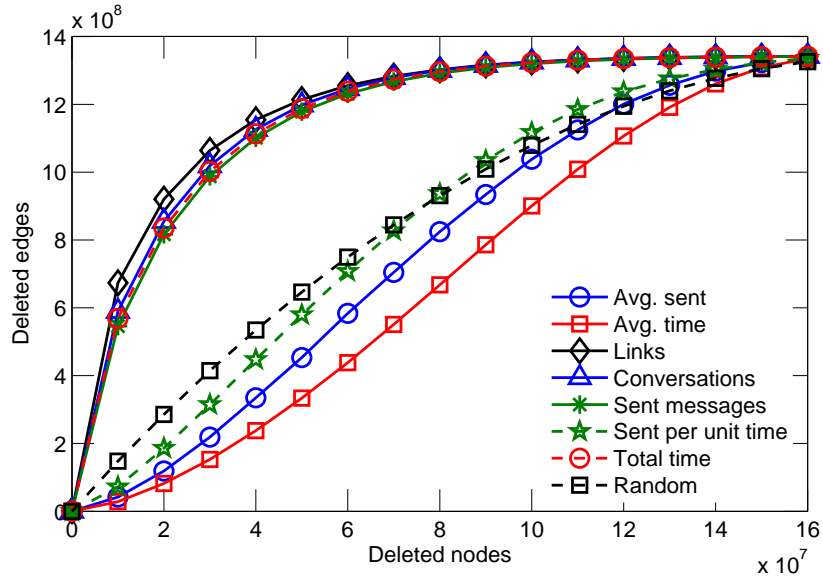}
  \caption{Number of removed edges as nodes are deleted by order of different measures of engagement.}
  \label{fig:commNet-delNodes}
\end{center}
\end{figure}

Figure~\ref{fig:commNet-delNodes} displays plots of the number of removed
edges from the network as  nodes are deleted. Similar to the relationships in
Figure~\ref{fig:commNet-delNodesWcc}, we found that deleting nodes by the
inverse number of edges removes edges the fastest. As in
Figure~\ref{fig:commNet-delNodes}, the same group of node ordering criteria
(number of conversations, total conversation time or number of sent messages)
removes edges from the networks as fast as the number of links criteria.
However, we find that random node removal removes edges in a linear manner.
Edges are removed at a lower rate when deleting nodes by average time per
conversation, average numbers of sent messages, or numbers of sent messages
per unit time. We believe that these findings demonstrate that users with
long conversations and many messages per conversation tend to have smaller
degrees---even given the findings displayed in
Figure~\ref{fig:commNet-delNodesWcc}, where we saw that removing these users
is more effective for breaking the connectivity of the network than for
random node deletion. Figure~\ref{fig:commNet-delNodes} also shows that using
the average number of messages per conversation as a criterion removes edges
in the slowest manner. We believe that this makes sense intuitively: If users
invest similar amounts of time to interacting with others, then people with
short conversations will tend to converse with more people in a given amount
of time than users having long conversations.

\section{Conclusion}
\label{sec:conclusion}

We have reviewed a set of results stemming from the generation and analysis
of an anonymized dataset representing the communication patterns of all
people using a popular IM system.  The methods and findings highlight the
value of using a large IM network as a worldwide lens onto aggregate human
behavior.

We described the creation of the dataset, capturing high-level communication
activities and demographics in June 2006.  The core dataset contains more
than 30 billion conversations among 240 million people.  We discussed the
creation and analysis of a communication graph from the data containing 180
million nodes and 1.3 billion edges. The communication network is largest
social network analyzed to date.   The planetary-scale network allowed us to
explore dependencies among user demographics, communication characteristics,
and network structure. Working with such a massive dataset allowed us to test
hypotheses such as the average chain of separation among people across the
entire world.

We discovered that the graph is well connected, highly transitive, and
robust. We reviewed the influence of multiple factors on communication
frequency and duration. We found strong influences of homophily in
activities, where people with similar characteristics tend to communicate
more, with the exception of gender, where we found that cross-gender
conversations are both more frequent and of longer duration than
conversations with users of the same reported gender. We also examined the
path lengths and validated on a planetary scale earlier research that found
``6 degrees of separation'' among people.

We note that the sheer size of the data limits the kinds of analyses one can
perform. In some cases, a smaller random sample may avoid the challenges with
working with terabytes of data. However, it is known that sampling can
corrupt the structural properties of networks, such as the degree
distribution and the diameter of the graphs~\cite{stumpf05subnets}. Thus,
while sampling may be valuable for managing complexity of analyses, results
on network properties with partial data sets may be rendered unreliable.
Furthermore, we need to consider the full data set to reliably measure the
patterns of age and distance homophily in communications.

In other directions of research with the dataset, we have pursued the use of
machine learning and inference to learn predictive models that can forecast
such properties as communication frequencies and durations of conversations
among people as a function of the structural and demographic attributes of
conversants.  Our future directions for research include gaining an
understanding of the dynamics of the structure of the communication network
via a study of the evolution of the network over time.

We hope that our studies with Messenger data serves as an example of
directions in social science research, highlighting how communication systems
can provide insights about high-level patterns and relationships in human
communications without making incursions into the privacy of individuals.  We
hope that this first effort to understand a social network on a genuinely
planetary scale will embolden others to explore human behavior at large
scales.

\section*{Acknowledgments}

We thank Dan Liebling for help with generated world map plots,
and Dimitris Achlioptas and Susan Dumais for helpful suggestions.

\end{document}